\def\mode{1} 
\if 0\mode
    \documentclass[preprint,superscriptaddress,floatfix,12pt]{revtex4-2}
    \usepackage[left]{lineno}
    \linenumbers
\else
    \documentclass[aps,pra,twocolumn,superscriptaddress,floatfix]{revtex4-2}
\fi

\usepackage{environ}
\usepackage{amssymb,amsmath}
\usepackage[utf8x]{inputenc}
\usepackage[T1]{fontenc}
\usepackage{siunitx}
\usepackage[version=3]{mhchem}
\usepackage[dvipsnames]{xcolor}
\usepackage{xfrac}
\usepackage{tabularx}
\usepackage{multirow}
\usepackage[caption=false]{subfig}
\newcommand\captionof[1]{\def\@captype{#1}\caption}

\usepackage{natbib}
\usepackage{nameref}
\usepackage[unicode=true]{hyperref}
\hypersetup{colorlinks=false,
            pdfborder={0 0 0}}
\usepackage{graphicx}
\makeatletter
\def \figwidth{
    \if 0\mode
        0.5\textwidth
    \else
        \columnwidth
    \fi
}
\makeatother

\usepackage[normalem]{ulem}

\newcommand{\UHeis}{18.6(8)}

\newcommand{\UHF}{8.3(2)}
\newcommand{\Udoped}{8.0(3)}
\newcommand{\THF}{0.05_{-0.05}^{0.06}}
\newcommand{\TcutoffHF}{6}
\newcommand{\NumAtom}{340}

\makeatletter
\newwrite\remember@figures
\AtBeginDocument{\InputIfFileExists{\jobname.dft}{}{}\immediate\openout\remember@figures=\jobname.dft}
\AtEndDocument{\immediate\closeout\remember@figures}
\NewEnviron{dfigure}[1]{%
    \immediate\write\remember@figures{%
        \noexpand\rememberfigure{#1}{\unexpanded\expandafter{\BODY}}%
    }%
}
\NewEnviron{dfigure*}[1]{%
    \immediate\write\remember@figures{%
        \noexpand\rememberfiguretc{#1}{\unexpanded\expandafter{\BODY}}%
    }%
}
\newcommand{\placefigure}[2][tp]{%
    \csname remembered@figure@#2\endcsname{#1}%
}
\newcommand{\rememberfigure}[2]{%
    \global\@namedef{remembered@figure@#1}##1{%
        \begin{figure}[##1]#2\end{figure}%
    }%
}
\newcommand{\rememberfiguretc}[2]{%
    \global\@namedef{remembered@figure@#1}##1{%
        \begin{figure*}[##1]#2\end{figure*}%
    }%
}
\makeatother

\begin{document}
\title{A neutral-atom Hubbard quantum simulator in the cryogenic regime}

\newcommand{\harvard}{Department of Physics, Harvard University, 17 Oxford St., Cambridge, MA 02138, USA}
\newcommand{\flatiron}{Center for Computational Quantum Physics, Flatiron Institute, 162 5th Avenue, New York, NY 10010, USA}
\author{Muqing~Xu}
\author{Lev~Haldar~Kendrick}
\author{Anant~Kale}
\author{Youqi~Gang}
\affiliation{\harvard}
\author{Chunhan~Feng}
\author{Shiwei~Zhang}
\affiliation{\flatiron}
\author{Aaron~W.~Young}
\author{Martin~Lebrat}
\author{Markus~Greiner}
\affiliation{\harvard}

\date{\today}

\begin{abstract}
Ultracold fermionic atoms in optical lattices offer pristine realizations of Hubbard models \cite{esslinger_fermi-hubbard_2010}, which are fundamental to modern condensed matter physics \cite{lee_doping_2006, arovas_hubbard_2022}.
Despite significant advancements \cite{tarruell_quantum_2018,gross_quantum_2021,bohrdt_exploration_2021}, the accessible temperatures in these optical lattice material analogs are still too high to address many open problems \cite{mckay_cooling_2011, altman_quantum_2021, qin_hubbard_2022, daley_practical_2022}. 
Here, we demonstrate a several-fold reduction in temperature \cite{mazurenko_cold-atom_2017, bohrdt_exploration_2021, xu2023frustration, chalopin_probing_2024}, bringing large-scale quantum simulations of the Hubbard model into an entirely new regime. 
This is accomplished by transforming a low entropy product state into strongly-correlated states of interest via dynamic control of the model parameters \cite{bernier_cooling_2009,lubasch_adiabatic_2011}, which is extremely challenging to simulate classically \cite{daley_practical_2022}. 
At half filling, the long-range antiferromagnetic order is close to saturated, leading to a temperature of $T/t=\THF$ based on comparisons to numerically exact simulations. 
Doped away from half-filling, it is exceedingly challenging to realize systematically accurate and predictive numerical simulations  \cite{qin_hubbard_2022}. 
Importantly, we are able to use quantum simulation to identify a new pathway for achieving similarly low temperatures with doping. 
This is confirmed by comparing short-range spin correlations to state-of-the-art, but approximate, constrained-path auxiliary field quantum Monte Carlo simulations \cite{he_finite-temperature_2019,qin_numerical_2017,xiao_temperature_2023}. 
Compared to the cuprates \cite{kivelson_how_2003,damascelli_angle-resolved_2003,lee_doping_2006}, the reported temperatures correspond to a reduction from far above to below room temperature, where physics such as the pseudogap and stripe phases may be expected \cite{kivelson_how_2003,zheng_stripe_2017,proust_remarkable_2019,xu_stripes_2022,arovas_hubbard_2022,simkovic_origin_2024}. 
Our work opens the door to quantum simulations that solve open questions in material science, develop synergies with numerical methods and theoretical studies, and lead to discoveries of new physics \cite{altman_quantum_2021, daley_practical_2022}.

\end{abstract}

\maketitle
\subsection*{Introduction}
The Hubbard model is a paradigmatic description of strongly-correlated electrons that is central to our understanding of quantum materials \cite{arovas_hubbard_2022}. 
After decades of concentrated study, it is strongly believed that this model hosts a variety of exotic quantum states \cite{chowdhury_sachdev-ye-kitaev_2022, szasz_chiral_2020}, including the intriguing phases observed in the cuprate high-temperature superconductors \cite{lee_doping_2006,proust_remarkable_2019,arovas_hubbard_2022}.
Despite its apparent simplicity, solving the two-dimensional, square-lattice Hubbard model has proved an outstanding challenge.
In recent years, building on decades of algorithmic development, combined use of the most advanced numerical methods has established certain aspects of the model's properties \cite{leblanc_solutions_2015,zheng_stripe_2017,qin_absence_2020,schafer_tracking_2021,xu_stripes_2022,iv_two-dimensional_2022,simkovic_origin_2024}.
Nevertheless, a full description of the low-temperature phase diagram remains beyond the reach of theoretical tools and computational methods \cite{qin_hubbard_2022}.

Its broad applicability and numerical intractability make the Hubbard model an ideal candidate for quantum simulation \cite{altman_quantum_2021,daley_practical_2022}. 
In particular, ultracold fermionic atoms in optical lattices can
natively implement the requisite fermionic degrees of freedom, and serve as a pristine model system for strongly-correlated electrons moving and interacting in a crystalline lattice \cite{esslinger_fermi-hubbard_2010}. 
The Hubbard model is described by the geometry of the lattice, a tunneling amplitude $t$ between neighboring sites, and a contact interaction energy $U$, which can all be widely programmed in cold-atom systems. 
However, the enlarged length and time scales in cold-atom systems lead to very small energy scales, making it challenging to reach regimes corresponding to real materials at low temperatures \cite{mckay_cooling_2011}. 
The lowest reported experimental temperatures of $T = 0.25(2)t$~\cite{mazurenko_cold-atom_2017, bohrdt_exploration_2021, xu2023frustration, chalopin_probing_2024} (Fig.~\ref{fig:scheme}a) in two-dimensional Hubbard quantum simulators with intermediate interaction strengths of $U/t\simeq 8$ correspond to an effective temperature of approximately $700$~K in the cuprates by normalizing temperatures with superexchange energy $J=4t^2/U$ \cite{kivelson_how_2003,damascelli_angle-resolved_2003,lee_doping_2006} (see Methods). 
One needs to reach well below room temperature in order to investigate properties relevant to cuprates, including pseudogap behavior, stripe order, and unconventional superconductivity \cite{kivelson_how_2003,zheng_stripe_2017,proust_remarkable_2019,xu_stripes_2022,arovas_hubbard_2022,simkovic_origin_2024}. 
New cooling schemes are therefore needed for cold-atom-based platforms to offer competitive simulations of quantum materials.

\if 1\mode
\placefigure{f1}
\fi

\if 1\mode
\placefigure{f2}
\fi

\if 1\mode
\placefigure{f3}
\fi

In this work, we demonstrate a substantial reduction in the temperatures achievable in cold-atom-based quantum simulations of the Hubbard model on a two-dimensional square lattice. 
At half-filling, we achieve temperatures of $T = \THF\,t$ with $U/t\simeq 8$ in a large system comprised of $\simeq \NumAtom$ lattice sites, which corresponds to a several-fold improvement of the state of the art \cite{mazurenko_cold-atom_2017, bohrdt_exploration_2021, xu2023frustration, chalopin_probing_2024}. 
We compare the measured spin correlations, which extend out to long range, to exact numerical results, and find excellent agreement.

The behavior of the Hubbard model is not fully understood at finite doping, which has hindered the development of cooling techniques in this regime, even at the level of theoretical proposals.
Here, we experimentally propose and demonstrate a novel pathway to realize low temperatures with finite values of doping $\delta$ between $2\%$ and $21\%$.
Detailed comparisons with state-of-the-art approximate numerical computations~\cite{he_finite-temperature_2019,qin_numerical_2017,xiao_temperature_2023} indicate that our temperatures are $T \lesssim0.1\,t$, comparable to those achieved at half-filling. 
These temperatures correspond to a reduction to significantly below room temperature in cuprates.

Our scheme hinges on the efficient preparation of product states with extremely low entropies ~\cite{chiu_quantum_2018}. 
Seminal proposals suggested that these states can be adiabatically connected to strongly-correlated states of interest to reach low temperatures~\cite{popp_ground-state_2006,bernier_cooling_2009,lubasch_adiabatic_2011,lin_quantum_2019,nan_quantum_2021}. 
Despite experimental progress with bosonic systems \cite{simon_quantum_2011,yang_cooling_2020,dimitrova_many-body_2023},
practical implementations of these schemes for fermions have been challenging beyond proof-of-principle realizations at half-filling~\cite{murmann_two_2015,chiu_quantum_2018,spar_realization_2022,yan_two-dimensional_2022}. 
This holds especially true in large Hubbard systems, where finite-time, out-of-equilibrium dynamics are out of reach of numerical simulations \cite{brown_bad_2019,ji_coupling_2021}. 
Our work therefore showcases how quantum simulators can experimentally address such optimization problems.

\subsection*{Experimental scheme}
The spirit of our protocol is similar to most cooling cycles.
In a first generalized compression step, entropy is extracted from the system of interest to a reservoir in thermal contact by reducing the density of states of the system \cite{ho_squeezing_2009,bernier_cooling_2009,chiu_quantum_2018}. 
After the system is isolated from the reservoir, increasing the density of states by generalized expansion decreases the temperature.
In practice, our scheme involves initializing cold atoms in a low entropy band insulator (BI), and transforming the BI into a strongly-correlated state of interest (Fig.~\ref{fig:scheme}b) \cite{chiu_quantum_2018,spar_realization_2022}.

The BI can be prepared with extremely high fidelity because 
the chemical potential lies within the band gap featuring low density of states, which 
allows for efficient entropy redistribution to a metallic reservoir~\cite{ho_squeezing_2009,bernier_cooling_2009,mazurenko_cold-atom_2017,chiu_quantum_2018,yang_cooling_2020}. 
We load a spin-balanced mixture of fermionic $^6$Li atoms in the lowest two hyperfine states into an optical lattice (long-spacing lattice). 
We set the magnetic bias field at $550$G close to a broad Feshbach resonance to minimize interaction strengths for BI formation.
Programmable optical potentials created by two separate Digital Micromirror Devices (DMDs) first confine about $\NumAtom$ atoms into a BI covering $170$ sites, then isolate the BI from the reservoir.
We estimate a low initial entropy per particle of $s = 0.025(4)k_B$~\cite{chiu_quantum_2018} based on a measured singly-occupied site density of $n_s=0.7(2)\%$ in a central disk of $\simeq 110$ sites, although an accurate estimation is complicated by gradients at the edge of the confining potential due to finite optical resolution (see Methods).

Keeping the subsequent transformation slow relative to the relevant many-body timescales is crucial to realizing low final temperatures. This is particularly challenging as this protocol involves a substantial change in lattice filling from two in the BI to around one atom per site in strongly-correlated Hubbard systems. Such a massive change in the state typically involves very slow many-body timescales, which requires excellent local control of density~\cite{murmann_two_2015,chiu_quantum_2018,yang_cooling_2020,spar_realization_2022,yan_two-dimensional_2022}.
To address this challenge, we take advantage of an optical lattice whose geometry is dynamically adjustable~\cite{sebby-strabley_preparing_2007,greif_short-range_2013,xu2023frustration}.
In particular, we continuously split single lattice sites into double-wells, and subsequently connect these double-wells together into a square lattice, realizing a scheme similar to the one proposed in \cite{lubasch_adiabatic_2011}.
This doubles the number of sites at fixed total atom number (Fig.~\ref{fig:scheme}c,d and Extended Data Fig.~\ref{sfig:exp_info}), and naturally converts the BI into half-filled Fermi systems.
However, this doubling is not exact in the experiment due to variations of the optical potentials.
To precisely reach the target filling, our protocol further relies on tunable interactions and DMD potentials to ensure controlled preparation, isolation and expansion of the system of interest (see Methods).

\subsection*{Splitting an insulator}
Dynamically changing the lattice geometry leads to rich physics even at very large interactions of $U/t = \UHeis$, where the half-filled state resulting from splitting the BI is a Mott insulator with a large charge gap. Here, the Hubbard model simplifies to a spin model with antiferromagnetic Heisenberg coupling $J=4t^2/U$.
In the thermodynamic limit, this Heisenberg model is known to host a quantum phase transition between a 
disordered phase in the dimerized lattice, described as a product of isolated singlet states within double wells, and a N\'eel ordered phase in the square lattice with long-range antiferromagnetic correlations (Fig.~\ref{fig:half-filling}a)~\cite{katoh_phase_1993,matsumoto_ground-state_2001}. In our experiment, which is a finite-sized system, the many-body spin gap is expected to decrease monotonically during splitting, with a minimal value of $\Delta_H\propto J/L^2$ corresponding to the Anderson tower of states with broken symmetry \cite{anderson_approximate_1952}. These two lattice geometries are continuously connected by tuning the coupling between double wells, which preserves total spin $S_{\mathrm{tot}} = 0$ and enables efficient low-entropy state preparation~\cite{lubasch_adiabatic_2011}. 
We parametrize the transformation of lattice geometry by a single parameter $\alpha$ describing the ratios of tunneling amplitudes within and between double wells (see Fig.~\ref{fig:half-filling}a and Methods).

To track how spin order changes with lattice geometry, we experimentally measure the spin correlation function in the $z$ direction between sites at positions $\mathbf{r}$ and $\mathbf{r+d}$~\cite{xu2023frustration}:
\begin{align}
    \label{eq:correlation}
    C^{zz}_\mathbf{d}(\mathbf{r})&=\frac{1}{S^2}\left( \langle S^z_{\mathbf{r}}S^z_{\mathbf{r+d}} \rangle - \langle S^z_{\mathbf{r}}\rangle \langle S^z_{\mathbf{r+d}} \rangle \right).
\end{align}

In the $\alpha\rightarrow 0$ limit of disconnected dimers, which corresponds to the BI, we only detect saturated spin correlations between nearest-neighbor sites, $|\mathbf{d}|=1$ within the dimers and consistent with $0$ everywhere else (Fig.~\ref{fig:half-filling}b).
As these dimers are connected, spin correlations start to grow between the dimers at $\alpha\simeq0.7$ and become uniform and isotropic in the square lattice $\alpha = 1.0$. 

To probe spin correlations at longer bond distances $|\mathbf{d}|>1$, we obtain $C^{zz}_\mathbf{d}$ by averaging $C^{zz}_\mathbf{d}(\mathbf{r})$ over a Mott insulating region within a central disk of radius $r = 6$, and account for spatial inversion symmetry (see Methods). We confirm that spin correlations are antiferromagnetic and localized on the intra-dimer bonds in the dimerized lattice $\alpha = 0.3$, and that they extend to long range in the square lattice with $\alpha = 1.0$ (Fig.~\ref{fig:half-filling}c).

The nature of magnetic order during the lattice transformation is captured by the staggered magnetization $m^z$~\cite{mazurenko_cold-atom_2017}. We compute $(m^z)^2$ by averaging the sign-corrected spin correlations $(-)^{|d_x+d_y|} C^{zz}_\mathbf{d}$ up to a cutoff distance $d_{\mathrm{max}}$ (see Methods, Eq.~\ref{eq:staggered_magnetization}). We observe an increase of $(m^z)^2$ around $\alpha \simeq 0.7$ (Fig.~\ref{fig:half-filling}e), consistent with the range of critical values reported in previous numerical studies of dimerized Heisenberg models~\cite{katoh_phase_1993,matsumoto_ground-state_2001}. The experimental data is also consistent with quantum Monte Carlo simulations performed at a fixed temperature of $T = 0.5 J = 0.12 t$. 

We experimentally check ramp adiabaticity and confirm that the spin correlations have saturated by varying the ramp duration, which could be limited by heating. Surprisingly, we find the strengths of the DMD potentials confining the Mott insulator, which may lead to atom transport into and out of the central region, can strongly affect the final spin correlations, and therefore temperatures (see Methods and Extended Data Fig.~\ref{sfig:exp_data}). These measurements suggest charge transport may be crucial to temperature optimization~\cite{dolfi_minimizing_2015,soni_density_2016} in addition to spin dynamics of the idealized model.

\if 1\mode
\placefigure{f4}
\fi

\subsection*{Ultralow temperatures in the strongly-correlated regime}
At intermediate interaction strengths of $U/t \simeq 8$ (Fig.~\ref{fig:half-filling-hubbard}a), which are most relevant to cuprate physics, the Hubbard model remains an antiferromagnet at half-filling, and can thus be reached by the same adiabatic ramp as in the Heisenberg regime. 
However, due to the reduced interaction strength, the system has increased compressibility, which can lead to enhanced charge transport.
As a result, we find that the harmonic confinement produced by the lattice laser intensity profile results in density variation across the system after splitting. Thus, if the central density is to remain $n=1$, the edge of the system must be at $n<1$.
Since the entirety of the system began as a BI, this means atoms at the edge of the trap must be spilled out as the insulating gap is reduced during splitting. We achieve this by dynamically adjusting the strength of the DMD potentials (Fig.~\ref{fig:half-filling-hubbard}b).
During this step, we empirically find that it is helpful to allow for transport to occur at low interaction strengths, which could be related to the fact that a Fermi liquid is more compressible than a Mott insulator (Fig.~\ref{fig:doping}a)~\cite{soni_density_2016,brown_bad_2019,ji_coupling_2021}.
The timing and shape of the above ramps are extensively optimized based on the final temperature.

With the optimized experimental sequence, we observe antiferromagnetic spin correlations $C^{zz}_\mathbf{d}$ extending across the entire half-filling region with a radius $r=5$ for an interaction strength of $U/t = \UHF$ (Fig.~\ref{fig:half-filling-hubbard}c, see Methods).
We compare these experimentally measured correlations after sign-correction and azimuthal average $(-)^{|d_x+d_y|}C^{zz}_{d}$ to 
numerical results from simulations using finite-temperature determinant quantum Monte Carlo (DQMC) and ground-state auxiliary-field quantum Monte Carlo (AFQMC) methods, under open boundary conditions. At half-filling, the numerical results are in principle exact, although care must be taken to obtain accurate and unbiased spin correlations, especially at larger bond distances $d$ (see Methods).
We find the measured $(-)^{|d_x+d_y|}C^{zz}_{d}$ are close to saturation and consistent with numerical data at $T/t\in[0,0.1]$ (Fig.~\ref{fig:half-filling-hubbard}d).
These strong, long-ranged correlations translate to a narrow peak in the spin structure factor $S(\mathbf{q})$ at quasimomentum $\mathbf{q} = (\pi, \pi)$ (Fig.~\ref{fig:half-filling-hubbard}e), which is obtained as the Fourier transformation of $C^{zz}_{\mathbf{d}}$ (see Methods). Comparing numerical DQMC data to the experimental value of the squared staggered magnetization $(m^z)^2$ computed up to a cutoff $d_\text{max} = \TcutoffHF$ allows us to extract a temperature of $T/t=\THF$ (Fig.~\ref{fig:half-filling-hubbard}f and Extended Data Fig.~\ref{sfig:hf_temperature}). 

\subsection*{Doping cold antiferromagnets}
\label{sec:doping}
The most intriguing and poorly understood physics of the Hubbard model resides in the doped regime. At low temperatures of $T/t<0.2$, and with intermediate interactions of $U/t\simeq 8$, the doped Hubbard model is extremely challenging.
Unlike for half-filling, no numerically exact results for spin correlations $C^{zz}_d$ are available in this regime for sufficiently large system sizes, which have systematically explored parameters such as temperature and interaction.
Extending our cooling scheme to the above regime is therefore very valuable to advance our understanding, since quantum simulation can offer direct measurements, potentially adding a powerful new element in conjunction with approximate numerical simulations and analytic theory.

An outstanding challenge is that dopants must be coherently introduced and delocalized into the nearly half-filled system after splitting to avoid detrimental heating~\cite{dolfi_minimizing_2015,soni_density_2016,chiu_quantum_2018}.
To address this challenge, we must adapt our cooling scheme to reach a final state with a central density $n<1$.
We begin with a band insulator and then double the density of lattice sites, resulting in a central half-filled region.
Additional expansion of atoms out of this region is then required to achieve $n<1$.
In prior work, however, such expansion incurred substantial heating \cite{chiu_quantum_2018}, resulting in final temperatures of $T/t>0.4$.
In this work, we circumvent these challenges by performing the expansion in a shallow lattice, which increases the tunneling amplitudes and reduces interaction strengths, facilitating the efficient transport of particles.
At the same time, this decreases dissipative effects from the lattice light, reducing the background heating rate.

Our cooling scheme for doped systems is thus as follows (Fig.~\ref{fig:doping}a,b): after forming the BI, we reduce the lattice depth by $80\%$ and perform splitting. 
The large tunnelings and weak interactions transform the BI into a Fermi liquid.
We then expand the atoms by reducing the strength of the confining DMD potential, while ramping the magnetic bias field to its final value of $590$G.
The depth of the square lattice is then ramped up to its final value, reaching the strongly-correlated regime of $U/t = \Udoped$ (see Methods).
At the same time, the strengths of the DMD potentials are adjusted to minimize transport during lattice loading \cite{dolfi_minimizing_2015,soni_density_2016} (see Methods).

We report in Fig.~\ref{fig:doping}d the spin correlations $C^{zz}_\mathbf{d}$ averaged over a region-of-interest (ROI) of radius $r = 3$. We observe antiferromagnetic spin correlations extending over this entire region for hole dopings of up to $\delta=10\%$. To understand how spin correlations evolve with increasing doping, we plot the azimuthal average of the sign-corrected spin correlations $(-)^{|d_x+d_y|}C^{zz}_{d}$. 
Larger doping induces stronger relative suppression of spin correlations at longer bond distances, which reduces the range of antiferromagnetism. Interestingly, the absolute magnitude of correlation reduction seems insensitive to bond distances, which suggests doping decreases a long-range offset instead of affecting the short-range structure (Fig.~\ref{fig:doping}e).

We compare the experimentally measured spin correlations systematically against numerical results obtained with the constrained-path (CP) auxiliary field quantum Monte Carlo (AFQMC) method \cite{he_finite-temperature_2019,qin_numerical_2017,xiao_temperature_2023}, which is a state-of-the-art approximate numerical technique. The spin correlations predicted by CP-AFQMC grow monotonically as temperature decreases. For a hole doping range of $\delta\in [2\%, 21\%]$, we find good agreement between the experimentally measured nearest-neighbor spin correlations $C^{zz}_{1}$ and CP-AFQMC simulations at temperatures of $T/t\in[0,0.1]$. Simulations at $T/t=0.15$ exhibit significantly weaker correlations (Fig.~\ref{fig:doping}f).

\if 1\mode
\placefigure{f5}
\fi

Interestingly, for bond distances of $d>1$, although the agreement is good on the scale of the variation between modestly low temperatures ($T/t=0.25$) and the ground state, the experiment shows stronger correlations than predicted by CP-AFQMC. 
It is not quite at the level of the agreement found at half-filling between experimental data and numerically exact results for all bond distances.
We verify that our observations do not depend on the weak spatial variation of doping due to harmonic confinement by changing the radius $r$ of the experimental ROI, and further confirm that the system is at thermal equilibrium (see Extended Data Fig.~\ref{sfig:exp_data}).
Uncertainty in the calibration of experimental parameters (see Methods) is insufficient to explain the deviation, but other details in Hamiltonian difference (e.g. higher order corrections) could potentially contribute, especially at lower temperatures.
The discrepancy may also hint at an underestimation of long-range correlations by CP-AFQMC.
This would be consistent with data at elevated temperatures of $T/t=0.25,0.33$, where numerically exact DQMC results seem to be more consistent with experimentally measured spin correlations, and CP-AFQMC results show slightly lower correlations (see Methods and Extended Data Fig.~\ref{sfig:exp_qmc}).
These systematic comparisons are an example of the synergies that now exist between quantum simulations and numerics in the regime where obtaining exact numerical results are incredibly challenging.

Previous attempts \cite{chiu_quantum_2018} at adiabatic state engineering in the Hubbard model suffered from excessive heating during the transformation from BI to Hubbard systems.
Our scheme departs from such attempts in two ways: the use of a geometry tunable optical lattice enables us to halve the filling of the system without macroscopic density redistribution, and working at weak interactions in a shallow optical lattice minimizes heating effects during the remaining expansion.
To probe the the importance of the latter choice, we prepare a BI in a square lattice and expand it to a half-filled system in the same lattice by shaping the DMD potentials, and measure the temperature of the final state as a function of the expansion duration $\tau$ and the normalized lattice depth $\eta$ during expansion ($\eta=1$ indicates full depth).

To characterize the adiabaticity of the expansion, we scan the expansion duration at fixed depth $\eta=0.2$ (Fig.~\ref{fig:transport}a).
We observe a rapid increase in the temperature at short expansion durations, indicating $\tau = 20\mathrm{ms}\simeq 35\hbar/t$ as a critical timescale below which diabatic heating limits the temperature.
As the lattice depth is increased, the tunneling energy decreases, requiring longer expansion durations to remain adiabatic.
To probe the effects of lattice depth, we therefore measure the final temperature as a function of lattice depth at a fixed normalized expansion duration $\tau\simeq 60\hbar/t$, in which $t$ is computed from $\eta$ via a bandstructure calculation (see Methods).
As shown in Fig.~\ref{fig:transport}b, the final temperature rapidly rises when the lattice depth is increased beyond a critical value $\eta\simeq 0.6$, which cannot be due purely to single-particle adiabaticity effects.
Rather, we attribute it to a combination of interaction-induced heating during particle transport, which worsens as $U/t$ is increased, and dissipative effects induced by the lattice light and DMD light, which worsen due to the increase of $\eta$ and $\tau$.

The interaction-induced heating may be related to theoretical and experimental studies which have indicated that transport in strongly-correlated systems can experience an exponential slowdown~\cite{sensarma_lifetime_2010, schneider_fermionic_2012, ji_coupling_2021} and induce heating from density redistribution \cite{dolfi_minimizing_2015,soni_density_2016}.
We note that the interaction strength $U/t\simeq 1.5$ at the critical lattice depth indicated in Fig.~\ref{fig:transport}b is similar to the interaction at which a drastic increase in transport timescales was observed in \cite{schneider_fermionic_2012}.
These measurements highlight the importance of performing the expansion step in a shallow lattice, in addition to the use of a superlattice to double the density of sites.

\subsection*{Outlook}
In this work, we have demonstrated a substantial reduction in the temperatures achievable in large-scale quantum simulations of the Hubbard model, both with and without doping.
Through careful dynamical tuning of parameters in a programmable optical lattice, we are able to transform a trivial low entropy state into a cold, strongly-correlated quantum many-body state in equilibrium.
Critically, the dynamics involved in this transformation are exceedingly difficult to tackle via analytical techniques, or to simulate classically, and so must be optimized empirically using the quantum simulator itself.
The above approach is broadly applicable to situations where one can prepare low entropy product states with high fidelity, for example with optical tweezers~\cite{young_atomic_2024}.
These techniques can also be extended to other lattice geometries, including triangular and kagome lattices, which may host quantum spin liquids~\cite{szasz_chiral_2020}, and square lattices with diagonal tunneling $t'$, which may host unconventional superconductivity~\cite{xu_coexistence_2024}.

The temperatures and system sizes achieved in this work likely allow one to enter phases of the Hubbard model that have not yet been explored in cold atom quantum simulators, including phases involving charge order~\cite{proust_remarkable_2019,zheng_stripe_2017,xu_stripes_2022,arovas_hubbard_2022,bohrdt_exploration_2021}.
Recent advancements involving comparisons of different approximate numerical simulations in equilibrium~\cite{leblanc_solutions_2015, qin_absence_2020} have shed some light on these behaviors, including the interplay between the stripe and pseudogap phases and $d$-wave superconductivity~\cite{simkovic_origin_2024, xu_coexistence_2024}.
However, a complete microscopic understanding has remained elusive.
Quantum simulations in the appropriate regimes offer a unique opportunity to study features that are challenging to simulate classically, including spectral properties and real time dynamics.
Additionally, in equilibrium, quantum simulations provide a valuable extra data point that complements approximate numerical simulations by taking dramatically different assumptions and approximations.

While advanced numerical simulations have long informed the design and operation of cold atom-based quantum simulators, this work opens the possibility of a fruitful exchange, wherein results obtained via quantum simulation can also be used to develop and benchmark more efficient and accurate numerical techniques.
The benefits of combining classical and quantum tools go beyond simply serving as mutual benchmarks.
The success of the preparation scheme in this work suggests that, in the future, hybrid classical-quantum algorithms that use results obtained from a quantum simulator to optimize state preparation~\cite{cerezo_variational_2021, czischek_data-enhanced_2022} could provide significant advantages, leading to lower temperatures, and possibly a definitive understanding of the Hubbard model in and out of equilibrium.

\subsection*{Acknowledgements}
We thank Daniel Greif, Geoffrey Ji and Christie Chiu for early experimental contributions, Annabelle Bohrdt, Eugene Demler, Tilman Esslinger, Antoine Georges, Fabian Grusdt, Ehsan Khatami and Martin Zwierlein for insightful discussions, and Yuan-Yao He, Yiqi Yang and Zhou-Quan Wan for help with our calculations. 
We acknowledge support from the Gordon and Betty Moore Foundation, Grant No.~GBMF-11521;
National Science Foundation (NSF) Grants Nos.~PHY-1734011, OAC-1934598 and OAC-2118310;
ONR Grant No.~N00014-18-1-2863;
the Department of Energy, QSA Lawrence Berkeley Lab award No.~DE-AC02-05CH11231;
QuEra grant No.~A44440; ARO/AFOSR/ONR DURIP Grants Nos.~W911NF-20-1-0104 and W911NF-20-1-0163; 
ARO ELQ Award No.~W911NF2320219; 
the Flatiron Institute is a division of the Simons Foundation (C.F. and S.Z.);
the NSF Graduate Research Fellowship Program (L.H.K. and A.K.);
the AWS Generation Q Fund at the Harvard Quantum Initiative (Y.G.);
the Swiss National Science Foundation (M.L.);
the Intelligence Community Postdoctoral Research Fellowship Program at Harvard administered by Oak Ridge Institute for Science and Education (ORISE) through an interagency agreement between the U.S. Department of Energy and the Office of the Director of National Intelligence (ODNI) (A.W.Y.).

\begin{dfigure*}{f1}
    \centering
    \noindent
    \includegraphics[width=\linewidth]{"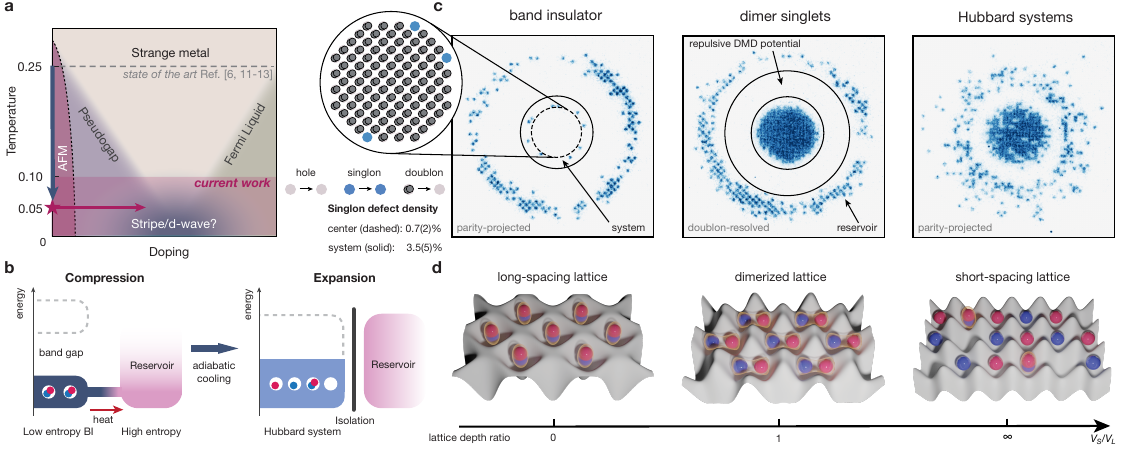"}
    \caption{\textbf{Cooling ultracold atoms by transforming a low entropy product state to strongly-correlated states.}
    \textbf{(a)}, Schematic of the 2D Hubbard phase diagram. The lowest temperatures reported in~\cite{mazurenko_cold-atom_2017, bohrdt_exploration_2021, xu2023frustration, chalopin_probing_2024} (gray dashed) and the estimated temperatures in the current work (red band) are marked.
    \textbf{(b)}, To reach lower entropy, we prepare a gapped band insulator (BI) in contact with a gapless metallic reservoir. Entropy flows from the BI to the reservoir~\cite{chiu_quantum_2018}. After isolating the two parts, the BI is transformed into a strongly-correlated state-of-interest by dynamically changing the lattice geometry, Hubbard parameters and density, while preserving the very low entropy of the original BI.
    \textbf{(c)}, The BI has a filling of two atoms per site, which appear as empty sites in a parity-projected fluorescence image (left), in contrast to the dilute reservoir. We then halve the lattice filling by doubling the number of lattice sites, thereby making the BI visible (center). The strongly-correlated state at the end of the preparation spans about $\NumAtom$ sites, where empty sites correspond to coherent doublon-hole pairs indicating low entropy (right). The central state-of-interest and reservoir are shaped by optical potentials programmed with Digital Micromirror Devices (DMDs, not shown).
    \textbf{(d)}, To double the number of sites, we continuously decrease the long-spacing lattice depth $V_L$ and increasing the short-spacing lattice depth $V_S$. This effectively splits doubly-occupied sites into dimers, which are then connected to form a square lattice.
    }
    \label{fig:scheme}
\end{dfigure*}

\begin{dfigure}{f2}
    \centering
    \noindent
    \includegraphics[width=\columnwidth]{"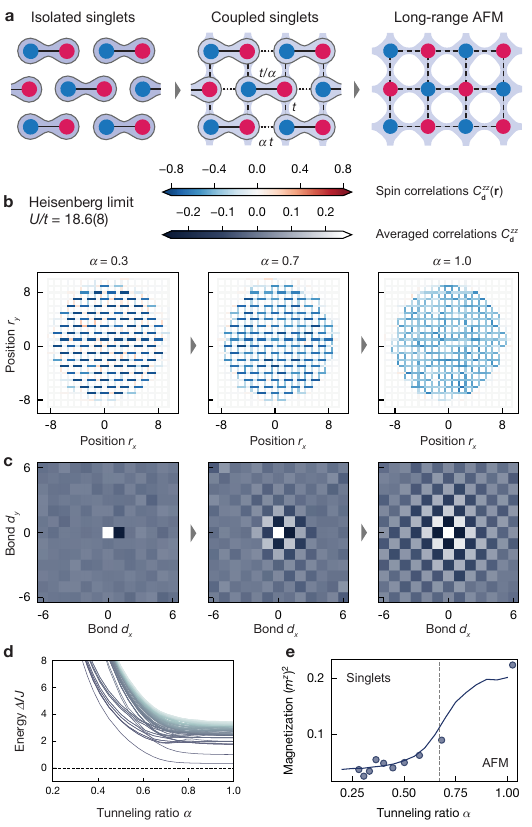"}
    \caption{\textbf{Splitting a band insulator into a Heisenberg antiferromagnet.}
    \textbf{(a)}, Similar to the long-spacing lattice limit, the ground state at half-filling on a disconnected dimerized lattice is a product state of singlets. This state is then adiabatically connected to a long-range antiferromagnet on a square lattice by coupling nearest dimers. 
    \textbf{(b)}, Spin correlations $C^{zz}_\mathbf{d}(\mathbf{r})$ between nearest-neighbors are localized on dimers when they are weakly coupled ($\alpha = 0.3$, left), and become uniform across all nearest-neighbor bonds in the square lattice limit ($\alpha = 1.0$, right). 
    \textbf{(c)}, Spin correlations as a function of bond displacement $C^{zz}_\mathbf{d}$, averaged over a $r=6$ region at half-filling at $U/t=\UHeis$. The range of the antiferromagnetic correlations starts to grow around $\alpha = 0.7$. 
    \textbf{(d)}, Energy levels of the Heisenberg model on dimerized lattices as coupling strengths are tuned, in units of the spin exchange coupling $J = 4t^2/U$ (see Methods).
    \textbf{(e)}, The measured staggered magnetization increases at couplings $\alpha > 0.7$ similar to the critical point of the quantum phase transition in the ground-state Heisenberg model \cite{matsumoto_ground-state_2001}. Solid lines: simulation of dimerized Heisenberg model at temperature $T/J = 0.5$.
    }
    \label{fig:half-filling}
\end{dfigure}

\begin{dfigure}{f3}
    \centering
    \noindent
    \includegraphics[width=\columnwidth]{"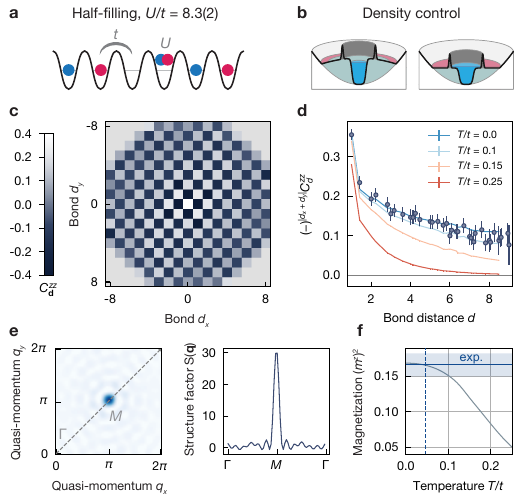"}
    \caption{\textbf{Cold Hubbard antiferromagnets in the strongly-correlated regime.}
    \textbf{(a)}, Intermediate interaction strengths, here $U/t = 8.3(2)$, lead to density fluctuations in addition to spin fluctuations at half-filling. 
    \textbf{(b)}, To achieve the target filling (here $n=1$), we adjust the confining potential created by DMDs to spill excess atoms. A repulsive wall isolates the cold region from the reservoir, while a volcano-shaped potential enables a spilling of excess atoms from the center.
    \textbf{(c)}, Spin correlations $C^{zz}_\mathbf{d}$ as a function of bond displacement $\mathbf{d}$, and averaged over a radius $r=5$ region, show nearly saturated long-range antiferromagnetic correlations. The error bars denote one s.e.m.
    \textbf{(d)}, Experimental data is compatible with numerically exact simulations (lines, DQMC and AFQMC) at temperatures $T/t < 0.1$ and interactions $U/t = 8$.
    \textbf{(e)}, Low temperatures yield a sharp peak of the spin structure factor $S(\textbf{q})$ at quasi-momentum $\textbf{q} = (\pi, \pi)$.
    \textbf{(f)}, We estimate temperature by computing the staggered magnetization with cutoff $d = \TcutoffHF$ on bond distance, for both DQMC--AFQMC data (interpolated to $U/t = \UHF$, line) and experimental data ($1\sigma$ confidence interval shown as shaded area). Their comparison yields a temperature $T/t = \THF$ (Methods).
    }
    \label{fig:half-filling-hubbard}
\end{dfigure}

\begin{dfigure*}{f4}
    \centering
    \noindent
    \includegraphics[width=\linewidth]{"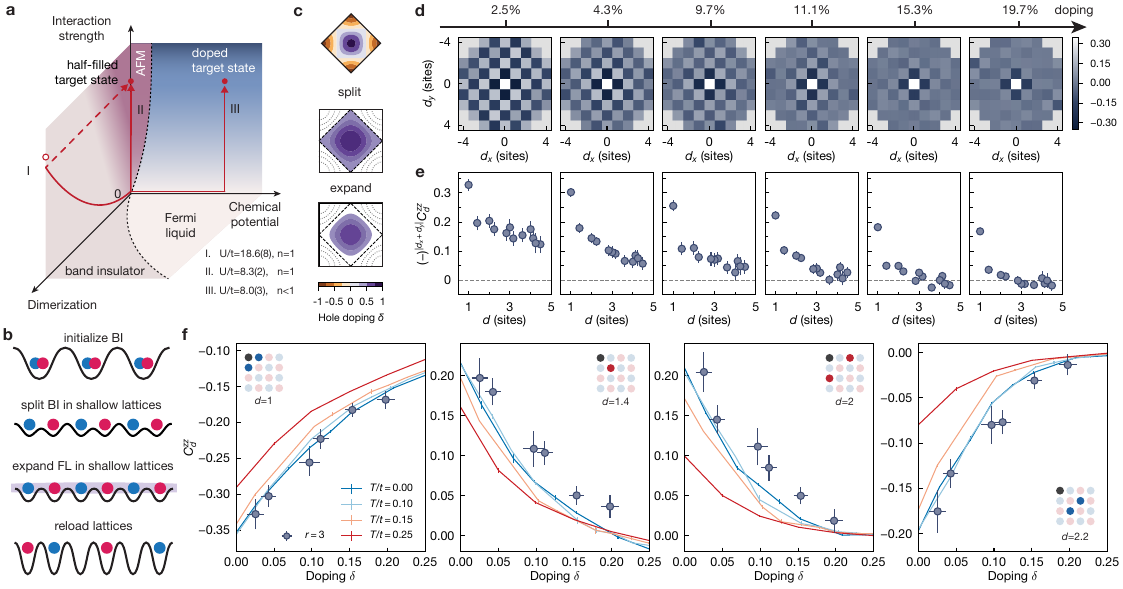"}
    \caption{\textbf{Hole-doping cold Fermi Hubbard systems.}
    \textbf{(a)},\textbf{(b)}, To coherently introduce hole dopants into the half-filled state after splitting the BI, we increase the kinetic energy and decrease the interaction strength by reducing the lattice depths out of the tight-binding regime. This prepares cold weakly interacting Fermi liquid that inherits the low initial entropy, which is adiabatically expanded to reach a given target density. Reloading the lattices and ramping up interactions create cold hole-doped Hubbard systems at $U/t=8.0(3)$ (III). Schemes of splitting the BI into a half-filled antiferromagnetic Mott insulator at strong interaction of $U/t=\UHeis$ (I) and intermediate interaction of $U/t=\UHF$ (II) are also marked.
    \textbf{(c)}, Doping scheme in quasimomentum space. Splitting the BI, which completely fills the Brillouin zone, into the square lattice doubles the number of sites and the size of the Brillouin zone. The population of quasimomentum states remain nearly unchanged. Expanding in real space then decreases the size of the Fermi surface and coherently introduces doping.
    \textbf{(d)} Spin correlations as a function of bond displacements $C^{zz}_{\mathbf{d}}$ measured in an ROI of $r=3$. The range and magnitude of the antiferromagnetic correlation decreases with increased doping. 
    \textbf{(e)} Azimuthal average of the sign-corrected spin correlations shown in \textbf{(d)}. At small dopings of $\delta\le 10\%$, the antiferromagnetism still remains long-ranged over the ROI. As we increase the number of hole dopants, the strengths of spin correlations at all distances are reduced by similar amounts.
    \textbf{(f)} Short-range spin correlations $C^{zz}_{d}$ at different distances $d=1,\sqrt{2},2,\sqrt{5}$ as we dope the system. The nearest-neighbor correlations at $d=1$ show quantitative agreement between experimental data and CP-AFQMC simulations at $U/t=8$ and $T/t\in[0,0.1]$ (solid line). At longer bond distances $d>1$, experimental data show stronger correlations than CP-AFQMC simulations. 
    The error bars denote one s.e.m.
    }
    \label{fig:doping}
\end{dfigure*}

\begin{dfigure}{f5}
    \centering
    \noindent
    \includegraphics[width=\columnwidth]{"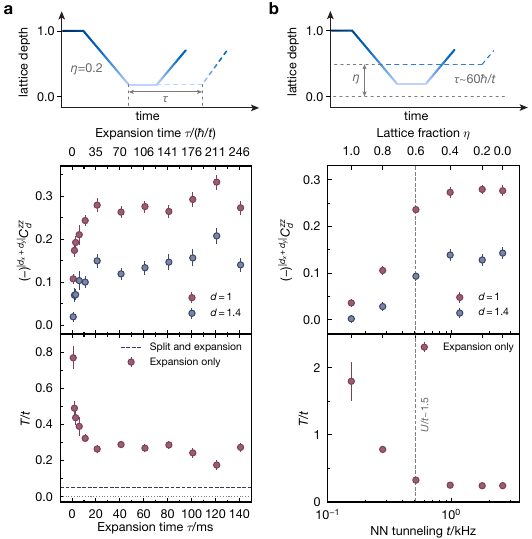"}
    \caption{\textbf{Adiabaticity of expansion.}
    \textbf{(a)} Adiabaticity of expansion from a BI to a to half-filled Hubbard system in the same lattice as a function of expansion duration $\tau$ at normalized lattice depth $\eta=0.2$. We find the ramp adiabaticity and therefore temperature improves quickly as the $\tau$ is increased from $1.76\hbar t$ ($1$ms), which starts to saturate at $\tau \simeq 35\hbar/t$ ($20$ms). 
    \textbf{(b)} Spin correlations and estimated temperatures after expansion of fermions for a fixed normalized duration measured by tunneling times $\tau \simeq 60\hbar/t$ (taking $\hbar=1$) at different lattice depths $\eta$. Expansion with $\eta\ge 0.8$ show strong heating which cannot be explained exclusively by adiabaticity. At critical lattice depth $\eta\simeq0.6$, the interaction strength $U/t\simeq 1.5$ is similar to the interaction at which a rapid increase in transport timescales was observed in \cite{schneider_fermionic_2012}. This indicates the heating may be largely dependent on a combination of interaction-induced effects and lattice heating.
    The error bars denote one s.e.m.
    }
    \label{fig:transport}
\end{dfigure}

\setcounter{figure}{0}
\renewcommand{\thefigure}{\arabic{figure}} 
\renewcommand{\figurename}{Extended Data Fig}
\renewcommand{\tablename}{Extended Data Table}

\clearpage
\section*{Methods}
\subsection{Lattice potential}
\label{s_sec:lattice}
In this work, the lattice potential is formed by three retro-reflected laser beams.
Two beams are overlapped and mode-matched, propagating in the $x$-direction, and are referred to as the $X$ and $\bar{X}$ beams respectively.
The third beam is propagating along the $y$-direction, orthogonal to the other two, and is referred to as the $Y$ beam.
As described in previous works \cite{xu2023frustration,lebrat2024observation}, the $X$ and $Y$ beams are phase stabilized to form an interfering lattice.
The frequency of the $\bar{X}$ beam is detuned by $\simeq 1.6$GHz from $X$ and $Y$ (Extended Data Fig.~\ref{sfig:exp_info}d,e).
The frequency offset is converted to a lattice phase shift of $\pi$ upon reflection from the retro-reflecting mirror, shifting the $\bar{X}$ lattice by half a site relative to the $X$ lattice.
This allows the $\bar{X}+Y$ lattice to split each unit cell in the $X+Y$ lattice symmetrically into two sites.
All three beams are reflected from a super-polished substrate in the $z$-direction at an angle of $\theta=69.2(1)^{\circ}$ before being retro-reflected \cite{xu2023frustration}, which forms a three-dimensional lattice potential.
The lattice potential in the $z$-direction provides confinement with a trapping frequency much larger than all relevant energy scales in the $xy$ plane lattices, and the tunneling along the $z$-direction is negligible during the experimental sequences.
This allows for a near-ideal realization of a two-dimensional system in the $xy$ plane after the atoms are selectively loaded into a single layer of the $z$ lattice.
The potential of the two-dimensional lattice can be written as

\begin{widetext}
    \begin{align}\label{eq:real_lattice_potential}
        V(x, y) = &-\frac{V_{x}\Bar{r}^{4}\left(1+\cos{(2\theta)}\right)}{4}\cos{(2k_{x}x)}+\frac{V_{\bar{x}}\Bar{r}^{4}\left(1+\cos{(2\theta)}\right)}{4}\cos{(2k_{x}x)} \\ \nonumber 
        &- \frac{V_{y}\Bar{r}^{4}\left(1+\cos{(2\theta)}\right)}{4}\cos{(2k_{y}y)} \\ \nonumber 
        &- \Bar{r}^{2}\cos^{2}{\theta}\frac{\sqrt{V_{x}V_{y}}}{4}\biggl(\cos{(k_{x}x-k_{y}y + \phi)} + \Bar{r}^{2}\cos{(k_{x}x+k_{y}y + \phi)} \\ \nonumber
        &+ \Bar{r}^{2}\cos{(k_{x}x+k_{y}y - \phi)} + \Bar{r}^{4}\cos{(k_{x}x-k_{y}y - \phi)}
        \biggr) \\ \nonumber
        &- \frac{V_{x}\Bar{r}^{2}(1+\Bar{r}^{4})\left(1+\cos{(2\theta)}\right)}{8} - \frac{V_{\bar{x}}\Bar{r}^{2}(1+\Bar{r}^{4})\left(1+\cos{(2\theta)}\right)}{8} \\ \nonumber
        &- \frac{V_{y}\Bar{r}^{2}(1+\Bar{r}^{4})\left(1+\cos{(2\theta)}\right)}{8}
    \end{align}
\end{widetext}

Here, $\Bar{r}=8.27(1)\%$ denotes the Fresnel loss present at the surface of the glass cell, $k_{x}=k_{y}=2\pi \sin{\theta} / \lambda$ are the lattice vectors in the $xy$ plane, and $\lambda = 1064$\,nm.
Owing to loss, the four interference terms cannot be combined except when the interference time phase $\phi$ between $X$ and $Y$ is $0$ or $\pi$.
Note that the lattice depths $V_{x,\bar{x},y}$ do not directly correspond to the lattice depths in an ideal retro-reflected square lattice, because of the finite incident angle in the $z$ direction and losses due to the presence of the glass cell.

\if 1\mode
\placefigure{exp_info}
\fi

\subsection{Experimental methods}
\label{s_sec:loading}
As in previous work \cite{xu2023frustration}, we prepare an ultracold, spin-balanced Fermi gas of $^6$Li atoms in the lowest two hyperfine states via evaporative cooling in a crossed optical dipole trap.
Spin balance is achieved through a microwave mixing process \cite{parsons2016probing} with a duration of $300$\,ms, ensuring the SU($2$) symmetry of the Fermi gas.
The ultracold $^6$Li atoms are then loaded from the optical dipole trap into the interfering (long-spacing) lattice.
The interfering lattice is a square lattice with $\sqrt{2}$ larger spacing than the non-interfering (short-spacing) lattice, and is formed by the actively phase-stabilized $X$ and $Y$ beams \cite{xu2023frustration} using equal intensity.
This initial lattice loading is performed in $100$\,ms, and the final lattice depths after loading are $V_X=V_Y=2.88(1)E_R$.
Here, $E_R=h^2/(8ma^2)=25.49(4)$\,kHz denotes the recoil energy and $a$ denotes the lattice spacing.
We set the magnetic field to $550$\,G, resulting in an $s$-wave scattering length of $a_s\simeq 84a_0$, and ensure that the applied field settles before lattice loading.
At the start of lattice loading, we turn on a blue-detuned confining potential formed by one of two DMDs, which we refer to as DMD0 and DMD1.
DMD0 is turned on over a duration of $60$\,ms (see Section.~\hyperref[s_sec:bandinsulator]{\textit{Characterize BI}}) and, together with the harmonic confinement provided by the lattice, controls the density profile of the atoms.
This redistributes the entropy in the system, creating a BI that is in thermal contact with a dilute metallic reservoir (see Section.~\hyperref[s_sec:bandinsulator]{\textit{Characterize BI}}).
Once the loading is complete, we turn on a second blue-detuned DMD potential projected using DMD1 in $5$\,ms~\cite{chiu_quantum_2018}, which acts as a wall to separate the BI from the reservoir in subsequent steps in the experiment.

\subsubsection{Experimental sequence for the half-filling data}
\label{s_sec:halffilling}
As described in \cite{lubasch_adiabatic_2011}, the phase transition from a BI to a half-filled Mott insulator with long-range antiferromagnetic correlations can be tuned by adjusting the ratio between inter- and intra-dimer tunnelling amplitudes.
In the Hubbard model, the intra-dimer tunneling $t_d$ is proportional to the band gap $\Delta_g$ in the BI limit.
The large band gap of $\Delta_g\simeq 100$\,kHz allows for a fast ramp of the $\bar{X}$ and $Y$ beams in $2$\,ms, while remaining adiabatic relative to motional energy scales.
The final $Y$ and $\bar{X}$ lattice depths are $9.29(3)E_R$ and $6.28(2)E_R$ respectively.
These values satisfy the relation $V_X+V_{\bar{X}}\approx V_Y$, ensuring that the harmonic confinement induced by the Gaussian intensity profile of the lattice beams is approximately rotationally symmetric.
Next, we ramp the depth of $X$ to $0.19(1)E_R$ in $15$\,ms, and simultaneously increase the depth of $\bar{X}$ to $9.06(3)E_R$,
This compensates for the decrease of $V_X$, and maintains $V_X+V_{\bar{X}}\approx V_Y$.
During this ramp, each site in the long-spacing lattice is split into a dimer, as shown in Extended Data Fig.~\ref{sfig:exp_info}d.
In quasimomentum space, this corresponds to reducing the band gap $\Delta_g$ between the ground band and the first excited band, while keeping the gaps from the ground band to higher bands nearly unchanged.
The ramp needs to be slower than the band gap $\Delta_g$ between the ground and first excited bands to be adiabatic.
This gap decreases with decreasing $X$ lattice depth.
We separate the ramp into two segments with a duration of $5$\,ms before and $10$\,ms after depths of $V_X=0.52(1)E_R, V_{\bar{X}}=8.74(3)E_R$.
The slower ramp at lower $X$ depths helps to accommodate the requirement of adiabaticity. 

The set point of $V_X$ is further reduced to $0.032E_R$, and $V_{\bar{X}}$ increased to its final value of $9.22(3)E_R$.
Note that there is a systematic uncertainty in $V_X$ below $0.1E_R$ due to residual leakage from the retro-reflected $\bar{X}$ lattice onto the $X$ lattice intensity regulation photodiode.
This causes an offset as large as $0.01E_R$ and decreases the actual $X$ lattice depth.
During the above ramp, the phase transition from spin singlets to antiferromagnetic order is expected to occur.
The ramp speed is $50$\,ms, which is a balance between adiabaticity and heating from the optical lattice.
Despite the presence of long range antiferromagnetic correlations, the resulting state still exhibits dimerized spin correlations due to the imbalanced inter- and intra-dimer tunneling along the $x$ direction (see Section.~\hyperref[s_sec:tunneling]{\textit{Calibration of tunneling and lattice parameters}}).
We further decrease the set point of the $X$ lattice over $30$\,ms to $0.016E_R$, which is the lowest depth that allows us to maintain phase stabilization between the $X$ and $Y$ lattices.
The depth of $V_X=0.016E_R$ is not sufficient to remove the dimerization of the tunneling, and so we further ramp the setpoint of the phase stabilization from $\phi=0$ to $\pi/2$.
Note that the contribution from phase noise to intensity noise is negligible at these low values of $V_X$.
Although this step is supposed to eliminate the interference lattice \cite{tarruell_creating_2012}, loss in the retro-reflected lattice beam (see Section.~\hyperref[s_sec:lattice]{\textit{Lattice potential}}) results in imperfect suppression of the interference term \cite{xu2024quantum}, and thus a potential offset between the $A$ and $B$ sublattices of the square lattice.
We use numerical simulations to confirm that this sublattice offset does not alter the relevant physics at half-filling (see Section.~\hyperref[s_sec:lattice_pot]{\textit{Lattice Potential Calibration}}).

To carry out the experiments shown in Fig.~\ref{fig:doping} (see Section.~\hyperref[s_sec:doped]{\textit{Experimental sequence for doped data}}) on the doped Hubbard model, we upgraded the apparatus with a non-reciprocal attenuator (see Section.~\hyperref[s_sec:isolator]{\textit{Non-reciprocal attenuator}}).
This allows us to decrease $V_X$ to a depth of $3.2(3)\times10^{-5}E_R$, corresponding to negligible tunneling dimerization.
With this upgrade, we can avoid ramping the phase stabilization setpoint $\phi$, and simply trigger the attenuator at the beginning of the dimerized lattice ramp.
As a verification, we repeat the experiments at half-filling with the attenuator while keeping $\phi=0$.
Under these conditions, we obtain a temperature of $T/t\simeq 0.1$, as shown in Extended Data Fig.~\ref{sfig:isolator}.
These measurements were performed without careful optimization, and obtained over a $12$-hour period without realignment.
The achieved temperature is consistent with those obtained without the attenuator for similar data taking durations without realignment.
We therefore attribute the slight increase in temperature compared to the data shown in Fig.~\ref{fig:half-filling-hubbard} to lattice drifts (see Section.~\hyperref[s_sec:alignment]{\textit{Effects of alignment}}).

In the short-spacing square lattice, the harmonic confinement imposed by the Gaussian intensity envelope is not compensated.
As a result, the final density profile is not flat except in a radius $r=5$ region near the lattice center at half-filling.
If all of the atoms in the initial BI are kept inside the DMD0 confining potential, the harmonic confinement will increase the chemical potential $\mu$.
According to DQMC simulations at the Hubbard interaction studied in most parts of this work of $U/t\simeq 8$, $|\mu|>t$ will lead to non-negligible doping $\delta>1\%$.
Therefore, harmonic confinement may introduce particle dopants into the system.
If the DMD0 potential had arbitrarily fine spatial resolution, allowing for an infinitely sharp wall, then as long as the DMD0 potential $V_D$ remains smaller than the band gap and $V_D\gg U$, the system would have a well-defined open boundary.
However, the resolution of the DMD0 potential is limited by the numerical aperture (NA=$0.7$) of the microscope objective, and the wavelength of $650$\,nm used to create the DMD potentials.
The excess atoms residing on the finite slope of the DMD0 potential will further increase the chemical potential $\mu$ of the central system as a function of the strength of the DMD0 potential, causing additional particle doping.
Such doping leads to charge transport through the system at strong Hubbard interaction, which we suspect leads to significant heating.
To minimize charge transport and heating, as described in the main text, the potential strength of DMD0 is scanned and optimized for each ramp.
We find the lowest temperatures are realized when the DMD0 potential is set to maintain $\mu\simeq 0$ in the lattice center, such that no transport occurs at this location throughout the entire ramp.
Close to the boundary of the DMD0 potential, the decreased local chemical potential causes excess atoms to flow out of the central region and form a secondary reservoir.
This could be beneficial for reducing the final temperature by realizing a secondary entropy redistribution step.
We start to ramp the magnetic field to its final value of $620$\,G after the dimerized lattice has mostly been formed, and where the density distribution matches that of the short-spacing square lattice. We find this is important to reach low temperatures, which we ascribe to interaction-induced heating in transport at large $U/t$ and to interaction-driven changes in the density profile.

The tunneling amplitudes for all datasets are summarized in Tab.~\ref{tab:dataset}.
Once the lattice and DMD0 ramp finishes, we immediately quench the lattice potential to $V_{\bar{X},Y}>60E_R$ within $50\mu$s to freeze the dynamics before imaging.
We confirmed that this quench is faster than doublon-hole dynamics and slower than band excitations.
The probed singlon density starts to increase once the quench time is slower than $100\mu$s due to the combination of virtual doublon-hole excitations, and starts to decrease once the quench time is faster than $20\mu$s due to excitation to higher bands.

\subsubsection{Experimental sequence for doped data}
\label{s_sec:doped}
The transport described in Section.~\hyperref[s_sec:halffilling]{\textit{Experimental sequence for half-filling data}} is only necessary due to imperfections of the potential, including finite optical resolution and harmonic confinement.
Splitting a BI in the long-spacing lattice naturally gives a half-filled state in the short-spacing lattice with $\mu\simeq 0$.
To reach a final state with finite doping, coherent redistribution of atom density across the entire system is required, which poses more challenges to suppress heating during atom transport.

After isolating the BI, we first reduce the lattice depth to $V_X=V_Y=0.50(1)E_R$ in $30$\,ms which increases the kinetic energy scale and further reduces the interaction strength before splitting into a Fermi liquid.
This depth corresponds to $\simeq 20\%$ of the original lattice depth, and is chosen to maximize the tunneling energy in the lowest band to enhance adiabaticity and reduce lattice-induced heating, while still maintaining a bandwidth smaller than the height of the DMD0 potential.
The latter constraint is important because otherwise the atoms would have enough kinetic energy to move out of the confining potential, leading to uncontrolled transport.
Another heuristic intuition is to maintain a finite band gap such that the Fermi surface before and after splitting are approximately matched.
In the non-interacting limit the quasimomenta of the atoms are conserved in a homogeneous lattice, and deformation of the occupation in quasimomentum space leads to diabatic transfer of atoms to excited states.
In an ideal tight-binding model with only nearest-neighbor tunneling, the full first Brillouin zone of the long-spacing lattice matches the Fermi surface of the half-filled first Brillouin zone of the short-spacing lattice (Fig.~\ref{fig:doping}c).
The approximate match between the occupied states before and after splitting may minimize the deformation of the Fermi surface, and therefore the heating process above.
A finite band gap also suppresses diabatic excitation into higher bands.
The reduced lattice depths before splitting weaken the confinement in the $z$-direction, potentially leading to tunneling in the $z$ direction.
To avoid this, we turn on a vertical lattice to allow the system to always be treated as two-dimensional.

\if 1\mode
\placefigure{isolator}
\fi

Unlike at half-filling, where the Mott insulating state is robust against potential variations, doped Hubbard systems are compressible and sensitive to potential offsets.
We therefore upgraded the lattices with a non-reciprocal attenuator that allows one to almost completely turn off the $X$ lattice, while maintaining phase stabilization between the $X$ and $Y$ lattices at $\phi=0$ (see Section.~\hyperref[s_sec:isolator]{\textit{Non-reciprocal attenuator}}).
An interference phase setpoint of $\phi=0$ ensures no potential offset, and therefore no density offset between the A and B sublattices.
We ramp on $\bar{X}$ to $0.50(1)E_R$, and ramp $X$ to $3.2(3)\times10^{-5}E_R$ in $50$\,ms.
Due to its low value, the $X$ depth is directly calibrated using a power meter (PM100D and S121C from \textit{Thorlabs}).
We trigger the attenuator at the beginning of this ramp to maintain phase stabilization while ramping $X$.
At the end of this ramp the lattice has negligible tunneling dimerization.
At this stage the splitting sequence is complete, and the first Brillouin zone doubled.

Expansion of the atom cloud is achieved by lowering the DMD0 potential in $50$\,ms, accompanied by a ramp of the magnetic field from $550$\,G to $590$\,G.
This expansion time is chosen to allow the magnetic field to settle.
Decreasing the DMD0 potential allows atoms to flow out of the confining potential, forming a dilute secondary reservoir confined by the DMD1 wall.
The DMD0 potential is chosen to adjust the central density to the desired value in the final state, which will minimize transport during lattice reloading.
Since expansion corresponds to a reduction of the Fermi momentum, a change in quasimomenta is necessary.
As a result, it is possible that increased scattering lengths due to the higher magnetic field may facilitate thermalization during expansion.

The prepared state is a weakly-interacting Fermi gas in equilibrium, which is similar to the state used to load the lattice in prior works~\cite{mazurenko_cold-atom_2017}, but at much lower temperature.
We then load the noninterfering lattice by ramping the $\bar{X}$ and $Y$ lattice depths up to $V_{\bar{X}}=11.0(1)E_R, V_Y=11.0(1)E_R$.
Here, the lattice depths are deeper than in the half-filling protocol to allow us to reach an interaction strength of $U/t\simeq 8$ at a lower magnetic field of $590$\,G (see Section.~\hyperref[s_sec:choice_lattice_field]{\textit{Choice of lattice depths and magnetic field}}).
In order to maintain the atom density profile, and minimize transport in the regime where $U/t$ is large, we lower the DMD0 potential and the DMD1 wall to accommodate reduced tunneling energies at higher lattice depths during the above ramp.
After lattice reloading, we freeze the dynamics within $50\mu$s before imaging.

\subsubsection{Potentials of DMD0}
\label{s_sec:dmd_potential}
We use an incoherent light source at $650$nm to illuminate the DMDs, which is blue detuned of the D1 and D2 transitions in Li, and forms a repulsive potential.
For experiments at half filling, the volcano-shaped potential created by DMD0 is composed of a paraboloid, which is chosen to compensate the harmonic confinement imposed by the lattice beams~\cite{mazurenko_cold-atom_2017,chiu_quantum_2018}, and a circular central cut out region that we will refer to as the crater (Extended Data Fig.~\ref{sfig:exp_info}).
The transition from the paraboloid to the crater is sharp (within $1$ pixel) on DMD0, which induces diffraction fringes on the potential due to the finite resolution of the imaging system.

To allow expansion for doped systems, we create the flattened volcano-shaped potential (see Section.~\hyperref[s_sec:expansion]{\textit{Adiabaticity of expansion after splitting}}) by applying a cut-off on the maximum amplitude of the volcano potential (Extended Data Fig.~\ref{sfig:exp_info}g).
This forms a flat ring-shaped region surrounding the crater.
As described in Section.~\hyperref[s_sec:bandinsulator]{\textit{Characterize BI}}, the flattened volcano-shaped potential may result in worse BI fidelity than without flattening. 
We choose to work with the largest flattened region for which we do not detect a reduction in BI fidelity.

\subsubsection{Non-reciprocal attenuator}
\label{s_sec:isolator}
As described in Section~\hyperref[s_sec:doped]{\textit{Experimental sequence for doped data}}, even when $\phi=0$, the intensity of the $X$ lattice needs to be reduced by more than in the half-filled case to avoid sublattice offsets.
This is challenging due to the reciprocal nature of most optical attenuators, including acousto-optical modulators or ND filters, which act on the beam both in the forwards direction (towards the atoms) and in the reverse direction (upon retro-reflection, returning to the detection photodiode).
A $50$dB attenuation of the $X$ lattice beam therefore leads to a $100$dB attenuation of the laser intensity on the phase stabilization photodiode, resulting in drastically decreased gain and signal-to-noise ratio, making it impossible to perform effective stabilization.
We solve this problem by introducing a non-reciprocal attenuator \cite{xu2024quantum} which applies variable and differing levels of attenuation to the forward and reverse beams.
This allows us to reduce the $X$ lattice depth by $5$ orders of magnitude while keeping the interference phase actively stabilized.
Under these conditions, we detect no tunneling dimerization or potential offset (see Section.~\hyperref[s_sec:verification]{\textit{Experimental verification}}).

\subsubsection{Effects of alignment}
\label{s_sec:alignment}
We find that the main source of instability in the experiment is the drift of the lattice positions (especially of the $\bar{X}$ lattice relative to the $X$ lattice).
This results in drifts in the lattice harmonic confinement, and corresponding shifts in the position of the peak atom density in the cloud.
The lattice position is affected by temperature and humidity fluctuations in the lab, and remains stable for $\sim0.5$ hour, which is the typical duration of a contiguous scan.
We observe drifts in lattice position on longer timescales which may lead to heating due to excess transport during the lattice ramps.
Empirically, we find the strongest effects of heating due to lattice drifts at half-filling.

For the doped data presented in Fig.~\ref{fig:doping}, we average over ROIs at the center of the trap with different radii, where the effects of harmonic confinement is minimized.
However, the lattice drifts shift the atom distribution relative to the DMD0 potential, which introduces a potential gradient within the ROI.
Therefore, for data presented in Fig.~\ref{fig:half-filling-hubbard},~\ref{fig:doping} we re-center the lattice beam positions every $\sim 1$ hour by maximizing the light back-coupled into the lattice optical fibers.

The drifts of the DMD potentials along the $x$ and $y$ directions are strongly suppressed by the magnification of the high NA objective used to project them.
We find that focus drifts along the $z$-direction occur over the course of weeks, which is convenient to correct.

\if 1\mode
\placefigure{exp_data}
\fi

\subsubsection{Choice of lattice depths and magnetic field}
\label{s_sec:choice_lattice_field}
The Hamiltonian of ultracold fermionic atoms moving and interacting in optical lattices naturally realizes the Hubbard model, with corrections in the form of beyond-nearest-neighbor tunneling, density-assisted tunneling, off-site interactions, and higher bands effects \cite{dutta_non-standard_2015}.
Shallow lattices are preferred from an experimental perspective, because at a fixed target value of $U/t$, larger tunneling energies make all energy scales higher.
Harmonic confinement is reduced in shallower lattices too, which, combined with the increased tunneling energies, results in a more homogeneous density distribution.
However, once the lattices are too shallow, longer-range tunneling grows appreciably, and the band gap decreases.
These effects lead to deviations from the Hubbard Hamiltonian and a breakdown of the tight-binding and single-band approximation.
As lattice depth increases the band gap increases, the Wannier function are more localized, a smaller scattering length $a_s$ is needed to achieve the same value of $U/t$, and the above corrections are exponentially suppressed compared to nearest-neighbor tunneling $t$ and onsite interaction $U$.
Deep lattices are therefore preferred from a theoretical perspective, and a tradeoff between experimental performance and asymptotic realization of the Hubbard model in an exact manner needs to be made.
In the non-interfering (short-spacing) lattice, in the half-filled case with $V_{\bar{X}}=9.22(3)E_R,V_Y=9.29(3)E_R$, which yields a radial band gap of $\Delta_{xy,g}\simeq 83$\,kHz and a vertical band gap $\Delta_{z,g}\simeq 42$\,kHz, we find the next-nearest-neighbor $\mathbf{d}=(2,0),(0,2)$ tunneling to be $t''\simeq 0.042t$.
In the doped case with $V_{\bar{X}}=V_Y=11.0(1)E_R$, which yields a radial band gap of $\Delta_{xy,g}\simeq 93$\,kHz and a vertical band gap $\Delta_{z,g}\simeq 47$\,kHz, we find the next-nearest-neighbor tunneling to be $t''\simeq 0.029t$.
Note that the radial lattice is nearly separable in the $x$ and $y$ directions, resulting in vanishing tunneling along directions other than $x$ or $y$.
Therefore, for the experimental parameters in this work, the single band, two-dimensional, and tight-binding approximations are well satisfied.
Note that the lattice depths provided here take Fresnel loss and the angle of polarization in the apparatus into account. The quoted depths are therefore higher than those of an idealized retro-reflected square lattice with the same band properties.

In practice, we find the maximum $s$-wave scattering length $a_s$ available to achieve the targeted Hubbard parameter $U/t$ sets the limit on the magnitude of tunneling amplitudes.
This is because shallower lattices yield larger tunnelings and smaller integrals of the Wannier functions, which requires increased $a_s$ to achieve a targeted $U/t$.
In the vicinity of the Feshbach resonance, the universal scaling of the fermion three-body recombination rate is $\kappa\propto a_s^6$ \cite{dincao_scattering_2005}.
Increasing the $s$-wave scattering length will eventually leads to excessive three-body loss and thus heating.
We find that if the atoms are kept in a lattice with depth $\simeq 10E_R$, as applies to the half-filling data, $a_s=512a_0$ at $620$\,G is the highest scattering length that does not lead to noticeable excess heating.
To achieve $U/t=8$, we set the lattice depths to $9.2E_R$.
To obtain the doped data, the lattices are ramped down to $0.5E_R$ for expansion.
Here, the lack of protection against three-body recombination from the lattice leads to significant heating at $620$\,G magnetic field.
We therefore choose to work at $a_s=294.5a_0$ at $590$\,G, and set the final lattice depth after reloading to $11E_R$.
Details of the lattice parameters are listed in Extended Data Table.~\ref{tab:dataset}.

\if 1\mode
\placefigure{qmc_data}
\fi

\subsection{Imaging procedure and fidelities}
\label{sec:imaging}
We perform site-resolved fluorescence imaging in the short-spacing square lattice as described in~\cite{parsons_site-resolved_2016}.
The fidelity of correctly determining the occupation of a lattice site is $F_i=99.4(6)\%$.

Imaging in the long-spacing lattice differs from imaging in the short-spacing lattice, and is described in~\cite{xu2023frustration}.
To image the BI with parity projection, we set the frequency detuning between $\bar{X}$ and $X$ to $850$\,MHz using an RF synthesizer, which ensures good overlap with the $X-Y$ and imaging lattices.
This allows for deterministic atom transfer between the physics and imaging lattices, despite the fact that the imaging lattice contains twice as many sites as the physics lattice.
As described in~\cite{xu2024quantum}, doublons are converted into molecules by ramping through a narrow Feshbach resonance at $543$\,G, and are subsequently lost due to light-assisted collisions.
We report a combined imaging fidelity, including detection fidelity and physics-imaging transfer fidelity, of $99\%$ \cite{lebrat2024observation}.

Imaging with full charge resolution in the long-spacing and dimer lattices is described in~\cite{lebrat2024observation}.
The dimer lattice is adiabatically connected to the long-spacing lattice before the gap between the ground band and the first excited band closes.
We set the frequency detuning of $\bar{X}$ relative to $X$ to $1552$\,MHz, which sets the position of the potential minimum of the $\bar{X}+Y$ lattice to be symmetric with respect to each unit cell in the long-spacing lattice formed by $X+Y$, and in the dimer lattice formed by $X+\bar{X}+Y$.
Each site in the long-spacing/dimer lattice is symmetrically split into two, with a negligible potential offset between the two minima of $\bar{X}+Y$.
In addition, we ramp the magnetic field to $610$\,G to generate strong on-site interactions between the atoms on doubly-occupied sites.
This facilitates adiabatic transfer of doubly-occupied sites in the long-spacing/dimer lattice to singly-occupied sites in the $\bar{X}+Y$ lattice.
We find the doublon detection fidelity to be $98\%$ after image reconstruction.

\subsection{Data analysis}
\label{s_sec:analysis}
In Fig.~\ref{fig:half-filling} (excluding subpanel b), \ref{fig:half-filling-hubbard} and Fig.~\ref{fig:doping}, two-point spin correlation functions are spatially averaged over all pairs of sites within a circular region of interest (ROI) centered on the atomic cloud.
Atomic density decreases away from the center as a result of the confining potential imposed by the lattice beams and the DMDs.
A large potential gradient would enhance the effective superexchange interaction $J$ for a site-to-site potential offset of $\Delta_V<U$, and suppress magnetic interactions with a kinetic origin~\cite{lebrat2024observation}.
We therefore chose a ROI radius of $r=6$ sites in Fig.~\ref{fig:half-filling}, $r=5$ sites in Fig.~\ref{fig:half-filling-hubbard}, and $r=3,4,5$ sites in Fig.~\ref{fig:doping}, in order to limit the potential variation.
This limits the variation of the radially averaged singlon density to $\simeq 2\%$ within the ROI.
The correlation maps for $r=4,5$ of the doped data are shown in Extended Data Fig.~\ref{sfig:exp_data}a,b.
For large dopings where only short-range antiferromagnetic correlations are present, the sign-corrected, radially averaged spin correlations $(-)^{|d_x+d_y|}C^{zz}_d$ may become negative at certain long bond distances.
This could be induced by harmonic confinement, or residual disorder of the lattice potentials, and will be explored in future works.

In bipartite dimerized lattices (Fig.~\ref{fig:half-filling}), we separately average the spin correlation function $\langle S^z_{\mathbf{r}} S^z_{\mathbf{r} + \mathbf{d}}\rangle_\mathcal{S}$ over reference sites $\mathbf{r} \in \mathcal{S}$ belonging to one of the two sublattices $\mathcal{S} = \mathcal{A}, \mathcal{B}$ and for all bond vectors $\mathbf{d}$.
Assuming inversion symmetry, we obtain and plot the sublattice-averaged spin correlation function $\langle S^z_{\mathbf{r}} S^z{\mathbf{r} + \mathbf{d}}\rangle = (\langle S^z_{\mathbf{r}} S^z_{\mathbf{r} + \mathbf{d}}\rangle_\mathcal{A} + \langle S^z_{\mathbf{r}} S^z_{\mathbf{r} - \mathbf{d}}\rangle_\mathcal{B})/2$.
All errorbars indicate the $1\sigma$ confidence interval obtained using bootstrap sampling across all experimental snapshots of a given dataset with $500$ randomly selected samples.
The number of experimental realizations for each dataset is reported in Extended Data Table.~\ref{tab:dataset}.

To convert the experimentally measured singlon density $n_{s,\mathrm{det}}$ to doping $\delta$ in Fig.~\ref{fig:doping}, we first correct for imaging fidelity to extract $n_{s,\mathrm{corrected}}=n_{s,\mathrm{det}}/F_i$.
We then use the doublon density $n_d$ as a function of density $n$ obtained from CP-AFQMC simulations at $T/t=0$ to reconstruct the doublon density as a function of singlon density $n_d(n_s)=n_d(n - 2n_d)$.
Finally, we compute the density $n_{\mathrm{exp}}=n_{s,\mathrm{corrected}}+2n_d(n_{s,\mathrm{corrected}})$ and doping $\delta_{\mathrm{exp}} = 1 - n_{\mathrm{exp}}$ of the experimental data.

In Extended Data Fig.~\ref{sfig:qmc_data}b, we plot the temperature dependence of the doublon density up to $n=0.85$.
Similar to half-filling, the temperature dependence is negligible compared to the statistical errors in our detected densities for $T/t<0.15$.
Although in these CP-AFQMC calculations, the $n_d$ results are not as accurate at higher densities of $n\in (0.85,1)$, interpolating between $n=0.85$ and half-filling suggests that the temperature dependence of the doublon density is still negligible for $T/t<0.15$.
This is supported by the fact that the energy scale associated with doublons is the interaction strength of $U\gg t\gg T$.
Therefore, using the ground state doublon density leads to negligible systematic errors in the reported quantities.

\if 1\mode
\placefigure{exp_qmc}
\fi

\subsection{Discrepancy between experimentally measured spin correlations and CP-AFQMC results}
\label{s_sec:dqmc_doped}
To better understand the discrepancy between the experimental measurements of beyond-nearest-neighbor spin correlations in the presence of doping and CP-AFQMC simulations in Fig.~\ref{fig:doping}, we explore a few possible sources of similar deviations.

As described in Section.~\hyperref[s_sec:calibration_U]{\textit{Calibration of Hubbard interaction $U$}}, we calibrate $U/t$ using the singlon density at half-filling, where $n_s$ reaches its maximum.
We also compare experimentally measured spin correlations as a function of the measured singlon density $n_s$ to $T = 0$ data from CP-AFQMC.
The singlon density is directly measured in the experiment, and does not involve converting $n_s$ to doping using numerical data.
We confirmed a miscalibration of $U/t$ seems insufficient to explain the observed discrepancy. However, higher-order corrections to the experimental Hamiltonian, including density-dependent terms, may lead to systematic deviations at finite doping from the calibration performed at half filling.
A complete comparison between experimental and numerical data requires a comprehensive characterization of the experimental Hamiltonian, and the development of calibration techniques that directly probe the doped regimes.

The constrained-path (CP) approximation used in the numerical simulations may also introduce systematic errors.
To explore this possibility, we compare experimental and CP-AFQMC data at elevated temperatures with numerically exact DQMC simulations.
When simulating the doped Hubbard model, the sign problem in DQMC leads to an exponential overhead with decreasing temperature.
Given accessible computational resources, we first perform DQMC at $T/t=0.25$ and $U/t=8$ in an $8\times 8$ system, and compare the results to those obtained with CP-AFQMC with the same parameters but in a $12\times 12$ system (Fig.~\ref{fig:doping}).
Both the DQMC and CP-AFQMC simulations use periodic boundary condition (PBC).
For these
DQMC simulations, we used the \textit{QUEST} package \cite{varney_quantum_2009} with a time step of $\delta \tau = 0.02$, $5000$ thermalization sweeps, and $200000$ measurement sweeps, which are repeated with random seeds for up to $4000$ runs for values of the density with severe sign problems.
The experimental data are obtained with $U/t=8.2(2)$, and using the standard loading sequence (without engineering the DMD potential) in Ref.~\cite{parsons_site-resolved_2016,xu2023frustration}.
The harmonic confinement due to the lattice intensity profile leads to a slow variation of the filling from the center to the edge of the trap.
We tune the atom number such that it reaches half-filling at the center of a region of radius $r=6$.
This allows us to perform accurate thermometry of the entire atom cloud by comparing to numerically exact simulation results at half-filling without the need of numerical results in the doped regime.
At larger distances from the trap center of $r>6$, the filling slowly decreases.
Radially binning the data allows us to probe spin-correlations as a function of varying singlon density and therefore doping at the calibrated temperatures. 

We focus on the short-range spin correlations $C^{zz}_d$ up to $d=\sqrt{5}$ in Extended Data Fig.~\ref{sfig:exp_qmc}a,b.
Given the short correlation lengths at this elevated temperature, and at finite doping, finite size effects are expected to be negligible.
Indeed, at a density of $n=0.995$ and chemical potential of $\mu/t=-1.0$, where the correlation lengths are expected to be longer than at larger doping, we find that the DQMC and CP-AFQMC simulations are in good agreement for all short-range correlations.
In addition, at $T/t=0.33$, we confirm that there is no difference between DQMC simulations performed in $8\times 8$ and $12\times 12$ lattices.

We find good agreement between experimental data, DQMC, and CP-AFQMC on the nearest-neighbor correlation $C^{zz}_1$ holds for all doping values at which we performed measurements or simulations.
However, for a doping range of $\delta\in [5\%,10\%]$ and 
bond distances $d=\sqrt{2},2,\sqrt{5}$, the results obtained from DQMC suggest stronger correlations, with which experimental data show good agreement (Extended Data Fig.~\ref{sfig:exp_qmc}a). This trend is consistent with the behavior shown in Fig.~\ref{fig:doping}.
The large statistical error bars in the DQMC data make it difficult to draw a definitive conclusion, especially because the estimated errors are themselves unreliable due to the vanishing average signs.

At even higher temperatures of $T/t=0.33$, the sign problem is less severe and we are able to compare DQMC and CP-AFQMC in a $12\times 12$ system.
We find smaller, but qualitatively similar, discrepancies where spin correlations $C^{zz}_{\sqrt{2}},C^{zz}_{2}$ computed from DQMC are stronger than CP-AFQMC for doping below $15 \%$ (Extended Data Fig.~\ref{sfig:exp_qmc}b).
Due in part to the reduced magnitude of the spin correlations, experimental data performed at this temperature appear statistically consistent with both numerical simulations.
Future work should be able to address this by using a better trial density matrix or an improved form of the self-consistent constraint in CP-AFQMC, and more systematic studies.

\subsection{Lattice Potential Calibration}
\label{s_sec:lattice_pot}

To measure the lattice trapping potential, we measure the density profile of a non-interacting spin polarized fermi gas loaded into the lattice, and, by taking the local density approximation (LDA), invert the non-interacting equation of state to obtain the local chemical potential.
The spin polarized Fermi gas is prepared by performing evaporative cooling with state 1 and 2 at $321$\,G, followed by a magnetic gradient-assisted spill-out of state 1 at $27$\,G, where state 2 is magnetically insensitive and therefore remains trapped.
Pauli exclusion in a spin-polarized Fermi gas prevents double occupancy (and associated parity projection during imaging), meaning that the measured density can be mapped unambiguously to a particular value of the chemical potential.
The uncertainty in the above calibration procedure is dominated by statistical errors, and is not strongly dependent on the assumed temperature of the Fermi gas.
We take a conservative estimate of $T/t = 0.5$ based on independent calibrations of the spin polarized Fermi gas, but the results do not change significantly when assuming a temperature in the range $T/t\in [0,1]$.

Using the above procedure, the harmonic confinement of doped data (Fig.~\ref{fig:doping}) is measured to be $V_H=0.0152(6)t/\mathrm{(site)}^2r^2$, where $r$ denotes the radial distance measured in sites from the lattice center.

The half-filled data at $U/t=8.3(2)$ (Fig.~\ref{fig:half-filling-hubbard}) was taken before the addition of the non-reciprocal attenuator, and so an offset between the A and B sublattices is present.
Via the procedure described above, we find that the difference in the mean local potential in the A and B sublattices is $\Delta\mu=0.75(3)t$.
This is in good agreement with the expected offset of $0.8(4)t$ given the geometry of the lattice.
The uncertainty in the expected offset is primarily due to uncertainty in the applied $X$ lattice depth at very low values.

To investigate the effects of sublattice offsets on spin correlations, we perform DQMC simulations at $T/t=0.15,0.3$ on an $8\times 8$ system, and CP-AFQMC simulations at $T/t=0$ on a $12\times 12$ system.
DQMC simulation at $T/t\ge 0.15$ show that an offset as large as $\Delta\mu=2t$ (defined as a symmetric sublattice offset about mean $\mu=0$) has little effect on spin correlations (Extended Data Fig.~\ref{sfig:qmc_data}a).
At $T/t=0$, CP-AFQMC shows no effects on the spin correlations for $\Delta\mu=t$, and a small decrease of spin correlations for $\Delta\mu=2t$.
This decrease is smaller than the statistical error in the experimental data, and so we conclude that sublattice offsets are not a concern for the above data set.

\subsection{Characterization of BI}
\label{s_sec:bandinsulator}
\subsubsection{Band insulator fidelity}
We characterize the fidelity of the BI using the singlon density $F_{BI}=1-n_s$, assuming that a perfect BI has a doublon on every site.
We can image the BI state with parity-projection in the long-spacing lattice, where doublons are lost due to light-assisted collisions (Fig.~\ref{fig:scheme}c, left) \cite{parsons_site-resolved_2016}.
Alternatively, we can directly image the doublon population by splitting each long-spacing site into two after freezing the dynamics to reconstruct the population with full density resolution (Fig.~\ref{fig:scheme}c, middle)~\cite{lebrat2024observation}.

Using density-resolved imaging, we measure the doublon density to be $n_d=98.2(5)\%$, the singlon density to be $n_s=1.8(5)\%$, and the hole density $n_h$ to be consistent with zero within a central ROI that doesn't include the boundary of the crater.
The ROI covers a circular region with a radius of $r=6$ sites in the long spacing lattice ($\simeq r=9$ in the short-spacing lattice), which is $1$ site ($2$ sites) smaller than the radius of the crater.
Note that given the fidelity of doublon detection in the experiment, the above numbers are consistent with a doublon fraction of unity.
The total number of atoms detected in the crater is $N_{r=10}=342(1)$.

The above measurements confirm that the empty sites appearing in the parity-projected images are doublons instead of holes. 
Given this observation, parity-projected imaging offers better sensitivity to the doublon population because it is immune to atom loss during fluorescence imaging.
The remaining errors in parity projected imaging of doublons include the fidelity of removing doublons through light-assisted collisions, and atoms hopping to a different two-dimensional layer during imaging.
We currently do not have an independent calibration of these errors, and so the measurements of the BI fidelity below constitute a lower bound.

Using parity-projected imaging, we find that most of the singly occupied sites occur at the edge of the crater.
We note that the crater is not aligned to the optical lattice in a site-resolved manner.
When combined with the finite resolution of the imaging system, this results in imperfectly controlled local potentials on the sites falling along the edge of the crater.
The equation of state on these sites may not favor doublons.
Additionally, the local density approximation may not hold in the presence of such abrupt potential variation.
We minimize the population of singlons on the edge by increasing the lattice depths and DMD0 potential strengths, such that the tunneling energy $t$ is small compared to the site-to-site potential difference at the crater edge.
We also set the magnetic field to $550$\,G to lower the Hubbard interaction $U$, which prevents the formation of a wide Mott plateau.
The detected singlon density is $n_{s,r=4}=0.5(3)\%$ in a central disk with $r=4$, and $n_{s,r=6}=0.7(2)\%$ in a disk with $r=6$.
Including the edge, the singlon population in the crater is $N_{r=10}=11.9(7)$.
However, the presence of singlons on the edge obscures the estimation of entropy in the BI because these singlons are not necessarily excitations.
On the other hand, because the density of states along the crater edge can be finite, these singlons can host a significant amount of entropy.
Taking this into account, we choose to define the BI fidelity $F_{BI}=1-n_{s,r=6}$ in the central region with $r = 6$ as a figure of merit to optimize the entropy redistribution procedure.

To optimize the initial loading of the BI, the DMD0 potential ramp is split into a linear ramp and a holding phase.
Once the lattice depths are high and the tunneling energy correspondingly small, the atoms may not redistribute for continued changes in the DMD0 potential (i.e. the ramp is not adiabatic).
We also limit the initial ramp rate to avoid creating band excitations in shallow lattices.

The volcano-shaped DMD0 potential allows the formation of a BI that is in thermal contact with a reservoir when the lattice is shallow, and isolated when the lattice is deep.
When using the flattened volcano-shaped potential, we find that the quality of the separation between system and reservoir deteriorates if the plateau surrounding the crater becomes too large, since some atoms may remain in the region with no potential gradient.

\subsubsection{Heating in the BI}
Heating processes in a BI are strongly suppressed due to the vanishing density of states, and the only active processes must excite atoms into higher bands.
Relevant heating mechanisms include incoherent light scattering from the optical lattice, intensity and position noise of the optical lattice, and background gas collision.
To quantify the loss, we hold the BI for a variable duration of up to $1600$\,ms and measure the total density $n$ using both parity-projected imaging and full charge-resolved imaging as a probe of BI fidelity.
In the parity-projected imaging we measure the defect singlon density in the long-spacing lattice.
In the full charge-resolved imaging we measure the density after splitting the doublons into singlons in the short-spacing lattice.
We find that the loss rate of density in the lattice center is $dn_{r=6}/dt=5.1(3)\%/\mathrm{s}$ using full charge-resolved imaging , and $dn_{r=6}/dt=4.8(4)\%/\mathrm{s}$ using parity-projected imaging, assuming that the detection of a singlon corresponds to the loss of a single particle (Extended Data Fig.~\ref{sfig:exp_data}c,d).
The zero-time intercept extracted from charge-resolved imaging is $n=1.983(3)$.
The intercept from parity-projected imaging is $n=0.008(4)$, which is consistent with the directly measured singlon density.
The difference between the parity-projected measurement and the charge-resolved measurement is consistent with the limitations imposed by imaging fidelity.
Given the loading time of $100$\,ms, the above loss rate is consistent with the measured BI fidelity of $F_{BI,r=6}=99.3(2)\%$.

\if 1\mode
\placefigure{experiment_verify}
\fi

\subsection{Calibration of Hubbard interaction $U$}
\label{s_sec:calibration_U}
Due to light-assisted collisions, in the short-spacing lattice we are only able to perform parity-projected density imaging, and therefore detect the density of singly-occupied sites $n_{s, \mathrm{det}}$ \cite{parsons_site-resolved_2016}.
We correct the detected singlon density by the imaging fidelity to estimate the actual singlon density $n_s$.
At half-filling (Fig.~\ref{fig:half-filling},\ref{fig:half-filling-hubbard}), the singlon density is a function of interaction and temperature: $n_s(U, T)$.
We rely on numerical simulations using DQMC to obtain the doublon density $n_d(U, T)$, and compute the singlon density as $n_s(U, T)=1-2n_d(U, T)$.
We find $n_d(U, T)$ is sensitive to interaction strengths, but has only a weak dependence on temperature for $0.15<T/t\le0.25$, and saturates for $T/t\le 0.15$ (Extended Data Fig.~\ref{sfig:qmc_data}c).
Using both the singlon density and spin correlations $C^{zz}_{\mathbf{d}}$, we estimate the possible ranges for interaction and temperature to be $8<U/t<9$ and $T/t\le0.15$.
The fact that $n_d$ is insensitive to temperature, and therefore effectively becomes a function of only interaction, allowing us to invert this equation of state to calibrate $U/t$ using only the measured singlon density $n_s$.
With $U/t$ calibrated, we can then estimate the temperature using spin correlations (see Section.~\hyperref[s_sec:calibration_T]{\textit{Calibration of Hubbard temperature $T$}}).

To obtain the doped data shown in Fig.~\ref{fig:doping}, the lattice depths and magnetic fields are slightly different from the ones used to obtain the half-filling data shown in Fig.~\ref{fig:half-filling-hubbard} (see Section.~\hyperref[s_sec:loading]{\textit{Experimental methods}}). Therefore, interaction strength $U/t$ needs to be re-calibrated for the doped systems. We adjust the amount of expansion to prepare a sample with half-filling $n=1$ in the center, which can be identified as the maximum of singlon density $n_s$ as filling $n$ is increased. We then apply the same calibration protocol described above to this half-filled state to obtain the Hubbard interaction $U/t$ given by the lattice and scattering length parameters used for the doped systems.

There is no available numerical simulation for the large $U/t$ Hubbard model at low temperatures, and so a different approach is required to calibrate the interaction strength $U/t$ for the data taken in the Heisenberg limit.
We first perform measurements at $580$\,G in the same lattice configuration as the final data, and with $T/t\le0.15$ such that the temperature dependence of the singlon density is negligible.
The resulting measured value of the singlon density is $n_s(580\mathrm{G})=0.898(7)$, corresponding to $U/t=8.5(4)$.
Next, using $a_s(620\mathrm{G})/a_s(580\mathrm{G})=512/235\approx 2.18$, we obtain $U/t=18.6(8)$.
If we ignore the temperature dependence of the singlon density and compare to numerical linked cluster expansion (NLCE) data at $T/t=0.2$ \cite{khatami_thermodynamics_2011}, our measured singlon density of $n_s(620\mathrm{G})=0.976(3)$ gives $19.4(1.7)$, which is consistent with the above measurement.

\subsection{Estimation of doping}

To convert singlon density $n_s$ to density $n$, we rely on numerical simulations using CP-AFQMC to obtain the doublon density $n_d(n, U, T)$ as a function of density, interaction strength, and temperature.
From this, we can compute the singlon density $n_s(n, U, T)=n-2n_d(n, U, T)$.
Similar to half-filling, the variation of doublon density reduces with temperature $T$, and becomes negligible for $T/t \le 0.15$ (Extended Data Fig.~\ref{sfig:qmc_data}b).

We plot the experimentally measured spin correlations as a function of singlon density $n_s$ together with numerically simulated spin correlations as a function of computed singlon density $n_s(n, U, T)$ in Extended Data Fig.~\ref{sfig:exp_data}f.
CP-AFQMC simulation data of $n_d$ is not shown in the density range $0.85<n<1$ due to poor convergence.
Based on comparisons of the nearest neighbor spin correlations, the data is consistent with temperatures of $T/t\le 0.15$ for all values of the doping where CP-AFQMC is performed.

Taken in combination, the above observations allow us to invert the equation of state and estimate the doping $\delta=1-n(n_s, U, 0)$ using numerical simulations performed at $T/t=0$, which have been studied systematically in previous works \cite{qin_numerical_2017}.

\subsection{Calibration of temperature $T$ at half-filling}
\label{s_sec:calibration_T}
For the half-filled data taken at $U/t = 8.3(2)$, good agreement between experimental and numerical data (Fig.~\ref{fig:half-filling-hubbard}) allows us to measure temperature using a physically motivated observable that aggregates information from spin correlations beyond nearest-neighbors.
Specifically, we consider the staggered magnetization $m^z$, which is defined as a sign-corrected average of the spin correlation function up to a cutoff $d$ in bond distance:
\begin{align}
   \label{eq:staggered_magnetization}
   (m^z)^2 &= \frac{1}{\mathcal{N}_{\Omega_d}} \sum_{(i, j), i^2 + j^2 \leq d^2} (-1)^{i+j} C^{zz}_{(i, j)}.
\end{align}
We estimate the value of $m^z_\text{exp}$, as well as the $1\sigma$ error $\delta m^z_\text{exp}$ in this estimate, from bootstrapping, and compute the expected value of $m^z_\text{num}$ as a function of temperature using the DQMC.

We check the effect of boundary conditions by obtaining DQMC numerical data on $12\times12$ and $16\times16$ square systems with open and periodic boundary conditions (Extended Data Fig.~\ref{sfig:hf_temperature}a).
Periodic boundary conditions tend to overestimate spin correlations $C^{zz}_d$ at long range, while the difference between $12\times12$ and $16\times16$ data is not substantial for $d\le6$.
This motivates our choice of to perform subsequent calculations using open boundary conditions and a $12\times12$ system.

Spin correlations are computed at $U/t = 8$ and $U/t = 9$ (Extended Data Fig.~\ref{sfig:hf_temperature}b), converted into magnetization $m^z_\text{num}$, and interpolated to produce a continuous function of temperature and interaction strength $m^z_\text{num} = f_U(T)$ (using cubic splines along $T$, and linear functions along $U$).
At a fixed value of $U$ based on experimental calibrations, we convert our knowledge of the true experimental magnetization, which we model as a Gaussian random variable $\tilde{m^z} \sim \mathcal{N}(m^z_\text{exp}, (\delta m^z_\text{exp})^2)$, into temperature by computing the probability distribution function of $\tilde{T} = g(\tilde{m^z})$, where $g$ is the inverse of the interpolated numerical data,
$g(m^z) = f_U^{-1}(m^z)$ if $m^z \leq f_U(0)$, and $g(m^z) = 0$ if $m^z > f_U(0)$.
We report the mean value and the $\pm1\sigma$ confidence interval of the temperature distribution.

Experimental and numerical magnetizations and their associated temperature estimates are shown in Extended Data Fig.~\ref{sfig:hf_temperature}c and d.
At the experimentally calibrated value of $U/t = 8.3$, increasing the bond cutoffs from $d = 3$ to $6$ (shown as vertical dashed lines in Extended Data Fig.~\ref{sfig:hf_temperature}a) does not change the inferred mean temperature of $T = 0.05$.
This indicates a good agreement between numerical and experimental data at all ranges, and confirms the calibrated interaction strength (Extended Data Fig.~\ref{sfig:hf_temperature}c).
With a cutoff $d = 3$ on bond distance, below which numerical data only weakly depends on boundary conditions (Extended Data Fig.~\ref{sfig:hf_temperature}a), the inferred mean value of $T/t$ varies by about 0.01 within the reported uncertainty on interaction strength $U/t = 8.3 \pm 0.2$, with again a weak variation of the confidence interval.
We report in the main text a temperature and asymetric errors of $T/t = 0.05_{-0.05}^{+0.06}$ obtained with a cutoff $d = 6$, and a single significant digit capturing the overall magnitude of the systematic errors related to interaction calibration and finite-size simulation effects.

\if 1\mode
\placefigure{hf_temperature}
\fi

\subsection{Calibration of tunneling and lattice parameters}
\label{s_sec:tunneling}
As described in Section.~\hyperref[s_sec:loading]{\textit{Experimental methods}}, the key step to ramp from a dimerized lattice to the short-spacing lattice is to ramp down the long-spacing lattice by reducing the $X$ lattice depth, and to ramp up the short-spacing lattice by increasing the $\bar{X}$ lattice depth.
Ramping the interference time phase $\phi$ from $0$ to $\pi/2$ eliminates tunneling dimerization, but introduces a potential offset between the A and B sublattices.
The non-reciprocal attenuator allows for $\phi$ to be stabilized over a much larger range of $X$ lattice depths, allowing us to suppress tunneling dimerization without introducing potential offsets.
We use the calibrated lattice depths and interference parameters \cite{xu2023frustration,lebrat2024observation} to numerically compute the band structure of the two-dimensional lattices.
The absolute amplitudes of tunnelings (in Hz) can then be computed by constructing maximally localized Wannier orbitals \cite{bissbort2012dynamical}, or by fitting the band structure with a tight-binding model.
We find that the two methods give consistent results for our lattice geometry (Extended Data Fig.~\ref{sfig:experiment_verify}a), and report the resulting tunneling amplitudes in Extended Data Table.~\ref{tab:dataset}.

\begin{table*}
    \begin{centering}
    \begin{tabular}{ | c | c | c | c | c | c | c | c | c |}
    \hline
    Dataset & $n_{s,\mathrm{det}} $ & $t_{d}$(Hz) & $t_{i}$(Hz) & $t_{y}$(Hz) & Field (G) & $U/t$ & Shots & Figures \\
    \hline
    DS1 & 0.894(8) & 1373(37) & 1373(22) & 1378(21) & 620 & 8.3(2) & 177 & \ref{fig:half-filling-hubbard} \\
    \hline
    DS2 & N/A & 854(27) & 854(15) & 877(17) & 620 & 18.6(8) & 1795 & \ref{fig:half-filling} \\
    \hline
    DS3 & 0.864(8) & 1045(19) & 1019(18) & 1056(19) & 590 & 8.0(3) & 201 & \ref{fig:doping} \\
    \hline
    DS4 & 0.855(7) & 1045(19) & 1019(18) & 1056(19) & 590 & 8.0(3) & 197 & \ref{fig:doping} \\
    \hline
    DS5 & 0.812(9) & 1045(19) & 1019(18) & 1056(19) & 590 & 8.0(3) & 140 & \ref{fig:doping} \\
    \hline
    DS6 & 0.800(9) & 1045(19) & 1019(18) & 1056(19) & 590 & 8.0(3) & 236 & \ref{fig:doping} \\
    \hline
    DS7 & 0.766(7) & 1045(19) & 1019(18) & 1056(19) & 590 & 8.0(3) & 366 & \ref{fig:doping} \\
    \hline
    DS8 & 0.744(7) & 1045(19) & 1019(18) & 1056(19) & 590 & 8.0(3) & 214 & \ref{fig:doping} \\
    \hline
    DS9 & N/A & 332(10) & N/A & 307(9) & 595 & 8.1(4) & 1700 & \ref{fig:transport} \\
    \hline
    DS10 & N/A & 332(10) & N/A & 307(9) & 595 & 8.1(4) & 1702 & \ref{fig:transport} \\
    \hline
    \end{tabular}
    \end{centering}
    \caption{Summary of experimental datasets. The detected singlon density $n_{s,\mathrm{det}}$ is averaged over a $r=5$ ROI for half-filling data (DS1), and a $r=3$ ROI for the doped data (DS3-8). In the DS2, we scan the dimerization parameter $\alpha$ in the Heisenberg limit with $U/t=18.6(8)$ to reach different final states. In the DS9 and DS10 we scan the lattice fraction and expansion time to probe heating under different conditions.}
    \label{tab:dataset}
\end{table*}

The tunneling dimerization is given by the interference term in the lattice potential, which scales as $\sqrt{V_X}$.
It is important to quantitatively determine the dependence of intra- and inter-dimer tunneling energies $t_d,t_i$, and the perpendicular tunneling energy $t_p$, on $V_X$.
With lattice depths $V_Y$ and $V_X+V_{\bar{X}}$ fixed, the variation of the different tunneling energies with $V_X$ are shown in Extended Data Fig.~\ref{sfig:experiment_verify}a.
At the end of the ramp $V_X=3.2(3)\times10^{-5}E_R$, and the tunneling energies are all balanced within systematic uncertainties.
To obtain the doped data, the $X$ is ramped down to $V_X=3.2(3)\times10^{-5}E_R$ in the shallow lattice condition, and kept the same for the rest of the experimental sequence.

\subsection{Experimental verification}
\label{s_sec:verification}
\subsubsection{Adiabaticity of splitting at strong interaction}
In the strong interaction limit $U/t=18.6(8)$, the adiabaticity is decided relative to the many-body spin gap for an ideal finite size system with sharp open boundaries and homogeneous potentials.
The gap separating the ground state and the Anderson tower of states scales as $1/L^2$ while the gap for spin wave excitations scales as $1/L$, where $L$ is the linear system size \cite{lubasch_adiabatic_2011, anderson_approximate_1952}.
In the absence of heating, slower ramp speed results in more adiabatic evolution.

Experimentally, as we increase the ramp duration from $20$\,ms to $80$\,ms, crossing the critical point predicted by previous numerical works \cite{matsumoto_ground-state_2001}, we find the spin correlations in the final state reach saturation and do not increase (Extended Data Fig.~\ref{sfig:experiment_verify}g).
We suspect that for longer ramp durations heating competes with the improvement of adiabaticity.
Indeed, holding during or after the ramp results in strong heating, which manifests as a reduction of spin correlations.
We also find that heating is related to the magnetic bias field.
When holding in the middle of the ramp, the heating is much stronger at $620$\,G, where we operate to achieve $U/t=18.6(8)$, than at $550$\,G, where the BI is prepared.
This hints at heating due to three-body processes, which are strongly enhanced as scattering length $a_s$ is increased.

The saturation of spin correlations with increasing ramp time is measured at the same DMD potential strengths as when the BI is formed.
Surprisingly, we find that with weaker DMD potentials, $20$\,ms ramps are no longer adiabatic, and result in reduced singlon density and spin correlations (Extended Data Fig.~\ref{sfig:experiment_verify}g).
This suggests that atoms are leaving the crater region.
However, we do not have a complete understanding of this dynamical process because the time-dependence of the magnetic field ramp also plays an important role.
Higher magnetic bias fields lead to stronger interaction strengths, which compete against the confining potential and may alter the physics of the transport process.

\subsubsection{Adiabaticity of splitting vs tunneling dimerization ramp}
In a homogeneous Hubbard system at half-filling, or in a Heisenberg spin system, the adiabaticity criterion is determined by the speed of the ramp compared to the energy gap of spin excitations \cite{lubasch_adiabatic_2011}.
In our case the relevant ramp is of the tunneling dimerization, which is controlled by the interference phase.
Experimentally, at $U/t=18.6(8)$, we find a ramp time longer than $5$\,ms is sufficiently adiabatic, and allows intra- and inter-dimer spin correlations to equalize.
Ramps with a duration of less than $1$\,ms are diabatic and result in clear deviations.
Interestingly, this ramp has no detectable effects on longer range correlations, which is likely decided by the rate of the previous ramp across the quantum critical point.
For longer ramp times, we find significant heating when the DMD0 potential is not optimized, leading to atom transport.
The coldest reported temperatures are therefore obtained by dynamically balancing the confinement of the DMD0 potential, the tunneling energy, and interaction strengths.
At $U/t=8.3(2)$, we find heating is much less problematic, and choose a ramp time of $25$\,ms to ensure adiabaticity.
A systemic study of the phase transition and the resulting requirements on adiabaticity would require proper flattening of the potential, and site-resolved preparation of the initial BI.

\subsubsection{Characterization of tunneling dimerization in doped systems}
The lattice phase stabilization is kept active between consecutive experimental sequences, and the interference time phase is maintained.
The optical lattice position is therefore tracked at the single-site level in fluorescence atom images.
This allows us to probe spin correlations in a bond-resolved manner (see Section.~\hyperref[s_sec:analysis]{\textit{Data analysis}}).
Different $y$, intra-, and inter-dimer tunneling energies $t_y,t_d,t_i$ lead to different superexchange interactions $J_y,J_d,J_i=4t_y^2/U,4t_d^2/U,4t_i^2/U$ on the respective bonds, and thus different spin correlations.
In Fig.~\ref{fig:half-filling}, we show how ramping the interference time phase $\phi=\pi/2$ eliminates the tunneling dimerization and correlation imbalance in the Heisenberg limit.
To obtain the doped data, 
at the end of the ramp to $V_X=3.2(3)\times10^{-5}E_R$ (using the non-reciprocal attenuator), we observe that the $y$, intra-, and inter-dimer spin correlations agree to within statistical errors (Extended Data Fig.~\ref{sfig:experiment_verify}d).
The density profile also shows no offset between the A and B sublattices, confirming the absence of potential offsets even with $\phi=0$ (Extended Data Fig.~\ref{sfig:experiment_verify}b,c).

Tunneling dimerization has striking effects on the doped Hubbard systems.
In contrast to the half-filled case, where long range correlations are negligibly affected by dimerization under the conditions $V_X=0.016E_R,\phi=0$, the doped systems only show correlations within the dimers.
This suggests that dimerization will lead to different physics in the doped Hubbard systems, and so it is crucial to remove this dimerization to observe the physics of the isotropic doped Hubbard model.

\subsubsection{Adiabaticity of expansion after splitting}
\label{s_sec:expansion}
Splitting a band insulating state naturally gives rise to a half-filled state.
To obtain doped Hubbard systems, we decrease the DMD0 potential such that the atoms can expand out of the crater.
The boundary of the expansion is set by the wall formed by the DMD1 potential.
Expansion in this ring-shaped region can be strongly affected by the shape of the DMD0 potential.
We find that a volcano-shaped DMD0 potential (Extended Data Fig.~\ref{sfig:exp_info}f) with a sharp, outward slope (where the potential decreases as distance from the center increases) optimizes the separation between the BI and the reservoir.
However, with this potential makes it more difficult to expand the atoms adiabatically.
The sharp slope acts as a potential barrier separating the crater and the region closer to the DMD1 wall potential, which forms two potential minima in the radial direction.
Intuitively this is similar to the broken symmetry state in a double-well system.
The initial BI state is elevated from the true ground state with a small gap given by the tunneling amplitude between the double well, which is exponentially suppressed with increasing strength of the potential barrier created by DMD0.
An alternative, classical, picture is that as the atoms flow out of the crater, they will accelerate downhill on the sharp slope, which is not a reversible process and therefore not adiabatic.
A bowl-shaped potential which increases until reaching the DMD1 wall as the radial distance from the center increases alleviates the above issues, but will trap additional atoms during the BI preparation leading to increased entropy of the initial state.
To balance these considerations, we choose the flattened volcano-shaped potential as a compromise (Extended Data Fig.~\ref{sfig:exp_info}g).
We confirmed the energy gap during expansion is similar to the case where the potential has an inward slope using exact diagonalization in one dimension.

Larger tunneling and smaller $U/t$ leads to more adiabatic expansion for fixed expansion duration, suggesting that shallow lattices are favorable. However, we find this does not hold true in the limit where the lattices are almost fully extinguished. We perform round-trip measurements of the BI fidelity by first decreasing the lattice depths to the values used in expansion, and then increasing the lattice depths back to those used to obtain the BI state. Excitations created during the ramp will manifest as singlon defects in the BI. We find a significant increase in the number of such defects when lattice depths are decreased to below $0.3E_R$ during expansion. As a result, we opt to use a lattice depth of $0.5E_R$ during expansion, which corresponds to $\simeq 20\%$ of the lattice depth used to load the BI.
We believe this behavior is due to band gaps becoming smaller than the tunneling energy, allowing diabatic excitations to higher bands.
After ramping to these conditions, we start the splitting procedure which merges the two lowest bands in the long-spacing lattice into a single band in the short-spacing lattice. 

We experimentally investigate the dynamics during expansion by varying the duration of the expansion step.
We measure the final spin correlations after reloading for the expansion times of $\tau_{exp}=1$\,ms, $15.5$\,ms and $30$\,ms.
No statistically significant variation is observed in these measurements (Extended Data Fig.~\ref{sfig:experiment_verify}e), suggesting that $1$\,ms is sufficient for the expansion to be adiabatic.
This is consistent with the large bandwidth and Fermi energy of $\simeq 10$\,kHz during expansion.
To further facilitate thermalization during expansion, and to allow the field to settle before reloading happens, we ramp the magnetic field from $550$\,G to $590$\,G during the expansion step, which increases the scattering length $a_s$ from $\simeq 86 a_0$ to $\simeq 295a_0$, and set the expansion time to a much longer duration of $50$\,ms.

\subsubsection{Adibaticity of expansion with fixed duration}

In Fig.~\ref{fig:transport} we show the adiabaticity of expansion in the long-spacing lattice with fixed duration in units of the tunneling time $\hbar/t$.
The temperatures are obtained by comparing the $d=1$ and $d=\sqrt{2}$ spin correlations $C^{zz}_d$ averaged in the central half-filled region to DQMC simulations.
The goal is to probe adiabaticity as a function of interaction strengths, independent of the absolute value of tunneling amplitudes.
Although expansion in deeper lattices may lead to more lattice heating due to the increased laser intensity and smaller tunneling amplitudes, little heating was observed in previous work \cite{lebrat2024observation} involving measurements performed at similar lattice depths and ramp times as for the deepest lattice explored in this figure ($\eta=1$).
The increased kinetic energy scales at shallower lattices may also reduce the effects of potential disorder, whose contribution to expansion adiabaticity requires future studies.
We also measured the adiabaticity of expansion with a fixed duration of $\tau=120$\,ms as shown in Extended Data Fig.~\ref{sfig:exp_data}e, which corresponds to $18.5\hbar/t$ for $\eta=1$.
We find similar heating as in Fig.~\ref{fig:transport}a despite the holding time being $\simeq 4$ times shorter.
This suggests that diabatic heating and dissipative heating are occurring on a similar scale.
According to Fig.~\ref{fig:transport}b, however, $120$ms should be slow enough to incur only $\simeq 0.2t$ increase in temperature, even at $\eta=1$ (it would correspond to roughly $4$ms in Fig.~\ref{fig:transport}).
This indicates that interactions likely play a role in inhibiting transport at $\tau=120$ms by placing more stringent requirements on adiabaticity.

\if 1\mode
\placefigure{thermal}
\fi

\subsubsection{Thermalization of splitting at half-filling}
To check whether the half-filled state in Fig.~\ref{fig:half-filling-hubbard} has equilibrated, we added a hold time of $\tau_h=0,10,20$\,ms after ramping into the final antiferromagnetic state.
During this measurement, the lattice alignment relative to the DMD potentials was not optimized, which may explain the higher resulting temperatures.
We find no change in the singlon density as a function of radial distance $r$ from the trapping potential center, or in the overall profile of the spin correlations as a function of bond distance $d$ (Extended Data Fig.~\ref{sfig:experiment_verify}h).
However, we observe a slight decrease of the nearest-neighbor correlation, which may be attributed to heating during the hold time.
This suggests that the ramp to split a BI into a half-filled system does indeed realize a state in thermal equilibrium.

\subsubsection{Thermalization during reloading}
After unloading the BI, splitting into a half-filled Fermi liquid, and expanding into a Fermi liquid with filling $n<1$, the sample is in a similar state to previous works \cite{parsons_site-resolved_2016,mazurenko_cold-atom_2017}, but at lower entropy.
We reload the lattice using linear ramp-up of the lattice depth in $100$\,ms, which is chosen to balance adiabaticity and lattice heating.
In addition, we experimentally confirmed the final quantum state is in equilibrium by holding for $\tau_h=0$\,ms, $20$\,ms and $40$\,ms, corresponding to $0\hbar/t$, $20\hbar/t$ and $40\hbar/t$. No statistically significant variation is observed for these hold times (Extended Data Fig.~\ref{sfig:experiment_verify}f).

Furthermore, in the doped system we can cross-check the thermalization throughout the sample by comparing temperature estimates obtained at different locations in the cloud.
As described in Section.~\hyperref[s_sec:analysis]{\textit{Data analysis}}, the temperatures shown in the main text are obtained in a ROI with radii $r=3,4,5$ in the center of the cloud using including only pairs of sites within the ROI (which we refer to as strict binning).
We can also obtain temperature estimates from the radially averaged spin correlations, although these estimates suffer from averaging over larger potential variations and different bin shapes.
Nevertheless, in Extended Data Fig.~\ref{sfig:thermalization} we reanalyze the data from Fig.~\ref{fig:doping}f to plot $C^{zz}_{\mathrm{bin}}(d)$ against the measured singlon density $n_{s,\mathrm{bin}}$. 
$C^{zz}_{\mathrm{bin}}(d)$ denotes the measured $zz$ spin correlation at bond distance $d$ for the corresponding bin.
Note that the above analysis does not apply strict binning, so only one of the two sites used to compute a given correlation function must be contained within the bin.
Although this leads to systematic deviations from the results obtained with strict binning
For example, the correlations with longer bond distance will be underestimated for the bin with the shortest distance since sites at lower densities are involved.
Nevertheless, this analysis offers a qualitative view that the atom cloud has thermalized across different densities.

\subsection{Exact diagonalization of BI splitting}
\label{s_sec:splitting_physics}
We use the \textit{QuSpin} package \cite{weinberg_quspin_2017} to compute the evolution of energy levels during splitting with exact diagonalization (ED).
For the Heisenberg model (Fig.~\ref{fig:half-filling}), we choose a system size of $6\times5$, with periodic a boundary condition in the long direction and twisted boundary condition in the short direction.
The twisted boundary condition is defined as connecting $(i, 5)\rightarrow((i + 1)\%6, 1)$.
The spin exchange coupling along the $y$ direction is fixed to $J=1$, and $\alpha=\sqrt{J_d/J}=\sqrt{J/J_i}$ are varied from $10$ to $1$.

\subsection{Numerical simulations at half filling}
\label{s_sec:dqmc_hf}
Our simulations at half-filling were performed with finite-temperature determinant quantum Monte Carlo (DQMC) and ground-state auxiliary-field quantum Monte Carlo (AFQMC) methods.
AFQMC and DQMC are numerically exact formulations
for studying quantum many-body systems.
The approach begins with the Trotter decomposition, which breaks the original imaginary-time propagator into smaller pieces.
To handle the on-site interaction, an auxiliary field is introduced via the Hubbard-Stratonovich (HS) transformation, transforming the interaction operator into a sum of one-body operators.
These steps allow the partition function to be expressed as a sum over the products of two Slater determinants --- one for spin-up electrons and one for spin-down electrons --- whose matrix elements can be computed analytically.
However, the product may take both positive and negative values, leading to a sign problem (except in specific cases, like the half-filled 2D repulsive Hubbard model, where particle-hole symmetry eliminates this issue).

In our simulations (see Fig.~\ref{fig:half-filling-hubbard}), finite-temperature results are obtained using the DQMC method with 7,000 thermalization sweeps, 200,000 measurement sweeps, and a time step of $\Delta \tau = 0.05$.
Ground-state data are generated with the AFQMC method, employing the Metropolis algorithm with force bias updates \cite{shi_ground-state_Metropolis_2015}, thousands of sweeps, and a time step of $\Delta \tau = 0.02$.
We verify that the Trotter error at these values of $\Delta \tau$ is smaller than the statistical error, ensuring the robustness of our results.
To mitigate ergodicity issues and fluctuations in spin correlations caused by the SU(2) symmetry breaking in the spin-$z$ HS transformation, we adopt the charge HS transformation for both the finite-temperature and ground-state calculations.
Additionally, the infinite variance problem \cite{shi_infinite-variance_2016} in spin correlation measurements is significantly reduced
when using the charge decomposition instead of the spin decomposition.

\subsection{Constrained-path auxiliary field quantum monte carlo simulation}
\label{s_sec:afqmc}
For doped systems, we employ both finite-temperature \cite{zhang_finite-temperature-1999,he_finite-temperature_2019} and ground-state \cite{zhang_constrained-1997} constrained-path auxiliary-field quantum Monte Carlo (CP-AFQMC) methods to compute the system's properties at zero and finite temperatures, respectively.
The finite-temperature CP-AFQMC method shares several of the building blocks 
from the standard DQMC method, including the use of Trotter decomposition and the Hubbard-Stratonovich transformation.
However, it introduces two key differences.
First, 
in systems suffering from the sign problem, 
an exact condition is derived for the paths in auxiliary-field space which cause the sign problem \cite{zhang_finite-temperature-1999}.
It is shown that only a small subset of paths in the auxiliary-field space contribute 
to the partition function.
The remaining paths are symmetrically 
distributed and 
cancel each other,
which adds noise to the computation of observables.
To address this, the CP-AFQMC method imposes a constraint, acting as a gauge condition, to isolate and sample the relevant paths.
This constraint is exact when the many-body Hamiltonian is used in the constraint.
Second, to implement the constraint and sample configurations efficiently, the algorithm employs a branching random walk with importance sampling to construct complete paths.
In practical applications, an effective Hartree-Fock Hamiltonian is used as a trial Hamiltonian to define the constraint~\cite{he_finite-temperature_2019}.
The parameters $U_{\rm eff}$ and $\beta_{\rm eff}$ are tuned so that the spin correlations obtained by the effective Hamiltonian closely match those of the many-body system\cite{xiao_temperature_2023}.

Ground-state CP-AFQMC \cite{zhang_constrained-1997} is a projection-based quantum Monte Carlo method that leverages the principle that the ground state can be accessed by applying an imaginary-time evolution operator to an initial wave function, provided the trial wave function has a nonzero overlap with the ground state.
Similar to the finite-temperature algorithm, ground-state CP-AFQMC also combines the Trotter decomposition with the Hubbard-Stratonovich transformation to reformulate the imaginary-time evolution as a stochastic process --- a random walk through the space of Slater determinants.
Ground-state properties are calculated as statistical averages over the configurations sampled during this random walk.
To address the fermion sign problem, an unrestricted Hartree-Fock solution is used as a trial wave function, introducing a constraint that ensures each Slater determinant in the random walk maintains a positive overlap with the trial wave function.
Extensive benchmarks in the literature~\cite{leblanc_solutions_2015,zheng_stripe_2017,qin_absence_2020,xu_stripes_2022} demonstrate that CP-AFQMC delivers high accuracy for studying ground-state and finite-temperature properties of quantum many-body systems.
Our ground-state AFQMC calculations followed similar procedures to Ref.~\cite{qin_numerical_2017} (although at half-filling we applied the charge decomposition as discussed earlier.)  
In our simulations (see Fig.~\ref{fig:doping}), the Trotter time step is typically 0.05 and 0.01 in finite temperature and ground state calculations, and we verify that the Trotter errors are limited.
The population of random walkers is around 5000 in these calculations.

\if 1\mode
\placefigure{hqmc}
\fi

\subsection{Quantum Monte Carlo for Heisenberg model}

Finite-temperature numerical simulations of the Heisenberg model in Fig.~\ref{fig:half-filling} and Extended Data Fig.~\ref{sfig:hqmc} are performed with the \texttt{SpinMonteCarlo.jl} package \cite{spinmontecarlo}, using a loop quantum Monte Carlo algorithm with 1024 warmup updates and 8192 measurement updates, over $16\times16$ dimerized-lattice unit cells (512 sites) with periodic boundary conditions.
Individual spin correlators $\langle S_z(\mathbf{r}) S_z(\mathbf{r}') \rangle$ are grouped by pairs of sites $(\mathbf{r}, \mathbf{r'})$ with the same bond vector $\mathbf{r}-\mathbf{r'}$, distinguishing pairs that belong to the same sublattice from those that do not.
These correlators are averaged at each measurement update, resulting in a relative statistical error on the order of $10^{-3}$.
The resulting correlation maps are shown in Extended Data Fig.~\ref{sfig:hqmc}a for representative values of the coupling parameter $\alpha$ at temperature $T/J = 0.5$.
We estimate that the systematic errors due to finite size effects are at most $5\times10^{-3}$ in the square lattice at this temperature, based on a comparison to a $32\times32$ system.

Staggered magnetization $m^z$ is obtained by averaging the sign-corrected spin correlations over bond distance (Extended Data Fig.~\ref{sfig:hqmc}b), in this case truncated to a square $13 \times 13$ region (Extended Data Fig.~\ref{sfig:hqmc}a), and matching the region of interest used in the analysis of experimental data.
We observe a distinct increase and temperature-dependence of $m^z$ at couplings above $\alpha = 0.65$, which we attribute to the existence of a quantum phase transition to an antiferromagnetic phase at zero temperature.
This critical value is comparable to ground-state Heisenberg simulations performed in similar dimerized lattice geometries \cite{matsumoto_ground-state_2001} --- note that the parameter $\alpha$ in our work is defined as a ratio of tunnel couplings, and is related to ratios of antiferromagnetic spin couplings in the equivalent Heisenberg model by a square root.

After normalising experimental data by the square spin moment $n_s^2$, where $n_s = 0.95$ is the averaged singlon density in this dataset, we observe good agreement between numerical Heisenberg data versus bond distance and experimental spin correlations $|\mathbf{r}-\mathbf{r'}|$ in the square lattice at $U/t = 18.6(8)$ (Extended Data Fig.~\ref{sfig:hqmc}c).
A least-squares fit of correlations up to $|\mathbf{r}-\mathbf{r'}| \leq 8$ results in an estimated temperature of $T/J = 0.458(3)$.

\subsection{Effective temperature in Hubbard systems compared to cuprates}
The single band Hubbard model can be related to the high-$T_c$ superconducting cuprates through a few steps of reduction and approximation.
A more realistic approximate model for the cuprates is the three-band Hubbard model, in which tunneling $t$ and interaction strength $U$ can be correspondingly related to energy scales in cuprates \cite{lee_doping_2006}.
Therefore, it is not straightforward to relate the temperatures in the one-band Hubbard model with doping to an effective temperature in cuprates.

However, at half-filling, the physics is well understood in the Hubbard model \cite{arovas_hubbard_2022} and cuprates \cite{kivelson_how_2003, damascelli_angle-resolved_2003, lee_doping_2006}, and can be described by an antiferromagnetic Heisenberg model.
In this regime, the characteristic energy scale is the exchange energy $J$, which can be readily probed in cuprates, and is $J=4t^2/U$ in the one-band Hubbard model.
This correspondence allows us to convert the temperature in the half-filled Hubbard model using $T/J$ to an effective temperature in cuprates.
Taking $J=125(5)$~meV in YBCO \cite{kivelson_how_2003,damascelli_angle-resolved_2003} and $U/t=8$, a temperature of $T/t=0.25$ corresponds to $\simeq 725$~K, and a temperature of $T/t=0.05$ corresponds to $\simeq 145$~K.

\subsection*{Data Availability}
The datasets generated and analysed during this study are available from the corresponding author on request. Source data are provided with this paper.

\subsection*{Code Availability}
The code used for the analysis are available from the corresonding author on request.

\subsection*{Acknowledgements}
We thank Daniel Greif, Geoffrey Ji and Christie Chiu for early experimental contributions, Annabelle Bohrdt, Eugene Demler, Tilman Esslinger, Antoine Georges, Fabian Grusdt, Ehsan Khatami and Martin Zwierlein for insightful discussions, and Yuan-Yao He, Yiqi Yang and Zhou-Quan Wan for help with our calculations. 
We acknowledge support from the Gordon and Betty Moore Foundation, Grant No.~GBMF-11521;
National Science Foundation (NSF) Grants Nos.~PHY-1734011, OAC-1934598 and OAC-2118310;
ONR Grant No.~N00014-18-1-2863;
the Department of Energy, QSA Lawrence Berkeley Lab award No.~DE-AC02-05CH11231;
QuEra grant No.~A44440; ARO/AFOSR/ONR DURIP Grants Nos.~W911NF-20-1-0104 and W911NF-20-1-0163; 
ARO ELQ Award No.~W911NF2320219; 
the Flatiron Institute is a division of the Simons Foundation (C.F. and S.Z.);
the NSF Graduate Research Fellowship Program (L.H.K. and A.K.);
the AWS Generation Q Fund at the Harvard Quantum Initiative (Y.G.);
the Swiss National Science Foundation (M.L.);
the Intelligence Community Postdoctoral Research Fellowship Program at Harvard administered by Oak Ridge Institute for Science and Education (ORISE) through an interagency agreement between the U.S. Department of Energy and the Office of the Director of National Intelligence (ODNI) (A.W.Y.).

\subsection*{Author Contributions}
M.X., L.H.K., A.K., Y.G., A.W.Y. and M.L. performed the experiment, collected and analysed data. C.F. and S.Z. performed the numerical DQMC, AFQMC and CP-AFQMC simulations, M.X. performed part of the numerical DQMC simulation in Methods, M.L. performed the Heisenberg model QMC simulations. M.G. supervised the study. All authors contributed to the interpretation of the results and production of the manuscript.

\subsection*{Competing Interests}
M.G. is co-founder and shareholder of QuEra Computing.

\begin{dfigure*}{exp_info}
    \centering
    \noindent
    \includegraphics[width=\linewidth]{"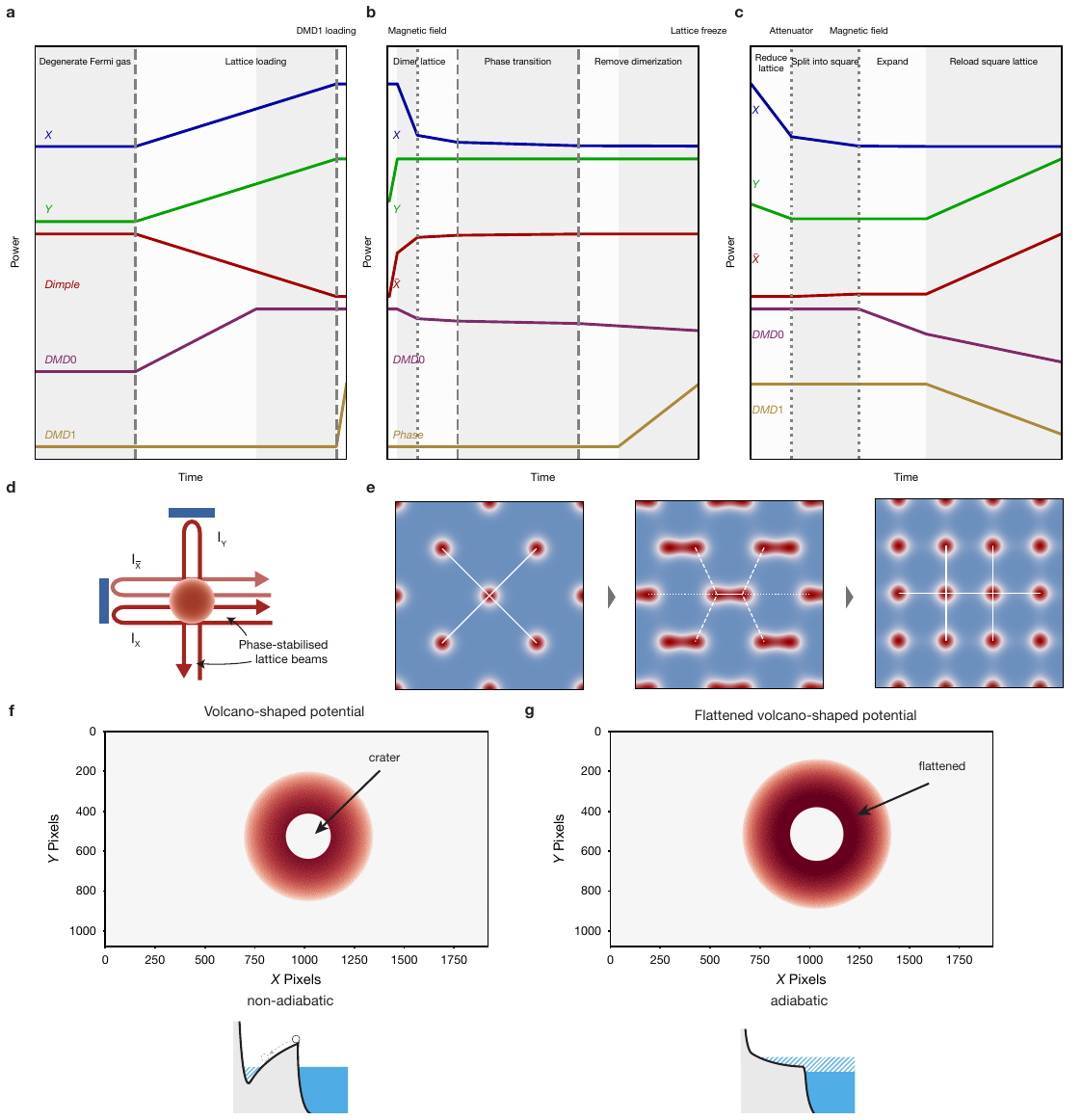"}
    \caption{\textbf{Schematic of the experimental details used in the cooling protocol.}
    \textbf{(a)}. The experimental sequence of the ramps used for the optical lattice and dipole potentials to initialize a BI with ultra-low entropy.
    \textbf{(b)}. The experimental sequence of the ramps used for the optical lattice and dipole potentials to split the BI into a cold antiferromagnet at half-filling in the $U/t=\UHF$ Hubbard model.
    \textbf{(c)}. The experimental sequence of the ramps used for the optical lattice and dipole potentials to split the BI and expand into a cold doped Hubbard system at $U/t=\UHF$. 
    \textbf{(d)}. Schematic of the lattice beams. Two orthogonal laser beams $X$ and $Y$, whose phase are interferometrically stabilized, are retroreflected to create an interference lattice. A frequency-offseted laser beam $\bar{X}$, co-propagating with $X$, are also retro-reflected to create a second lattice along the $x$ direction shifted by half a site. 
    \textbf{(e)}. As we decrease the intensity of the $X$ beam and increase the intensity of the $\bar{X}$ beam, the interfering long-spacing lattice is ramped down and the non-interfering short-spacing lattice is ramped up. The lattice geometry changes from a long-spacing square lattice through a dimerized lattice to a short-spacing square lattice.
    \textbf{(f)}. The volcano-shaped potential used in Fig.~\ref{fig:half-filling-hubbard}. A sharp, outward sloped potential facilitates preparation of the BI but prevents adiabatic expansion of the atom cloud.
    \textbf{(g)}. The flattened volcano-shaped potential used in Fig.~\ref{fig:doping}. Removing the potential barrier to form a flattened region allows adiabatic expansion. Each pixel of the DMD can be turned on or off. A continuous potential function is binarized using error diffusion algorithm \cite{mazurenko_cold-atom_2017}.
    }
    \label{sfig:exp_info}
\end{dfigure*}

\begin{dfigure*}{exp_data}
    \centering
    \noindent
    \includegraphics[width=\linewidth]{"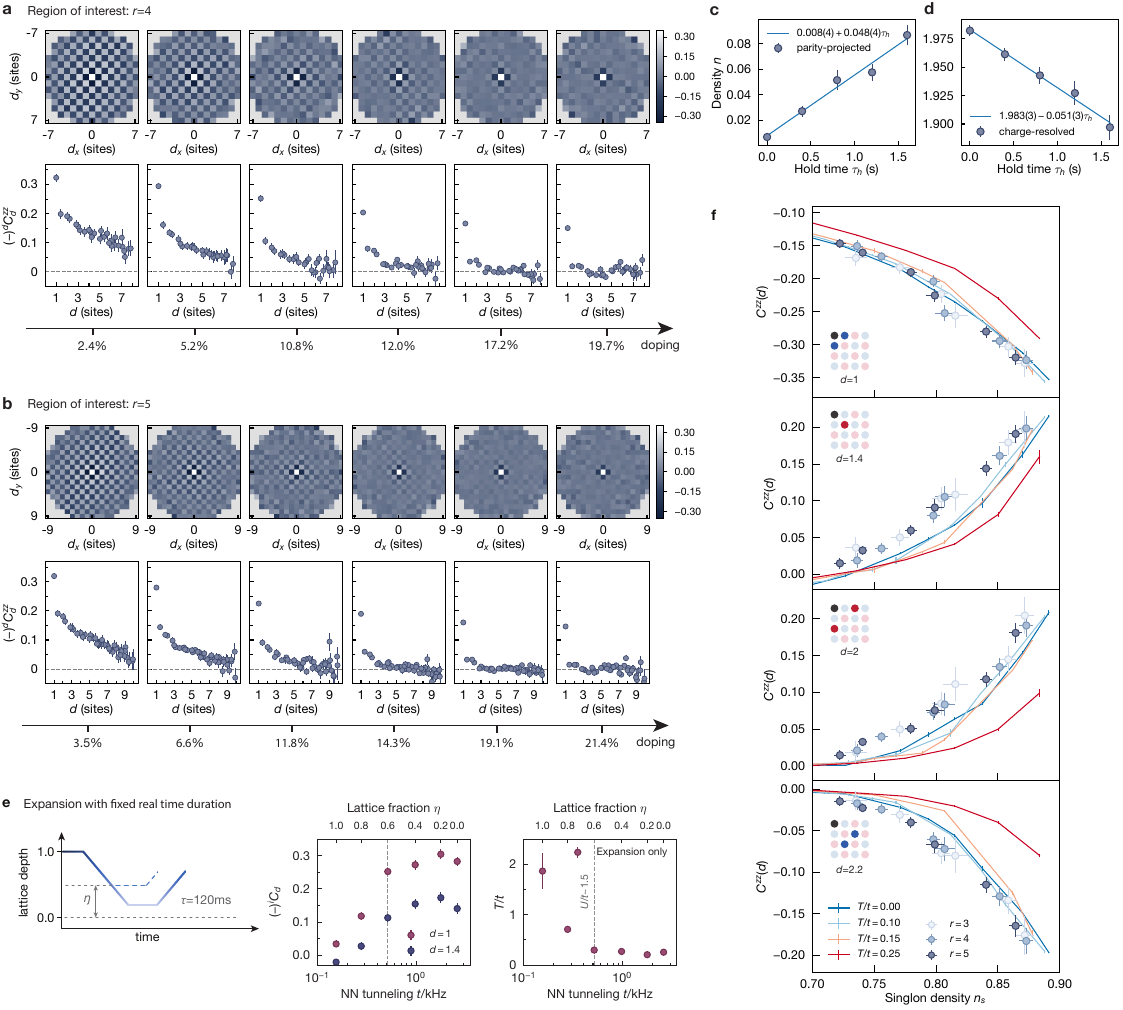"}
    \caption{\textbf{Supplementary experimental data.}
    \textbf{(a)}. Correlation map and the azimuthal average computed in an ROI of $r=4$.
    \textbf{(b)}. Correlation map and the azimuthal average computed in an ROI of $r=5$.
    \textbf{(c)}. Using parity-projected imaging, the defect singlon density measured in the long-spacing lattice as a function of holding time.
    \textbf{(d)}. Using full charge-resolved imaging, the density measured in the short-spacing lattice as a function of holding time. There is a factor of $2$ difference in the measured slope due to there are twice the number of sites.
    \textbf{(e)}. Here we allow the atom cloud to expand for a constant time $\tau_h=120$\,ms instead of constant time measured by tunneling times $\hbar/t$. The heating remain nearly the same despite the ramp time and therefore lattice heating is reduced by $\simeq 4$ times for the deepest lattice depth. The expansion time $\tau_h\simeq 15\hbar/t$ in the deepest lattice depth, which according to Fig.~\ref{fig:transport} should be sufficiently slow to cause negligible heating.
    \textbf{(f)}. Spin correlations $C^{zz}_d$ shown in Fig.~\ref{fig:doping} as a function of singlon density $n_s$. For experimental data we correct for imaging fidelity to obtain $n_{s, \mathrm{corrected}}$ as $n_s$. 
    }
    \label{sfig:exp_data}
\end{dfigure*}

\begin{dfigure*}{isolator}
    \centering
    \noindent
    \includegraphics[width=\linewidth]{"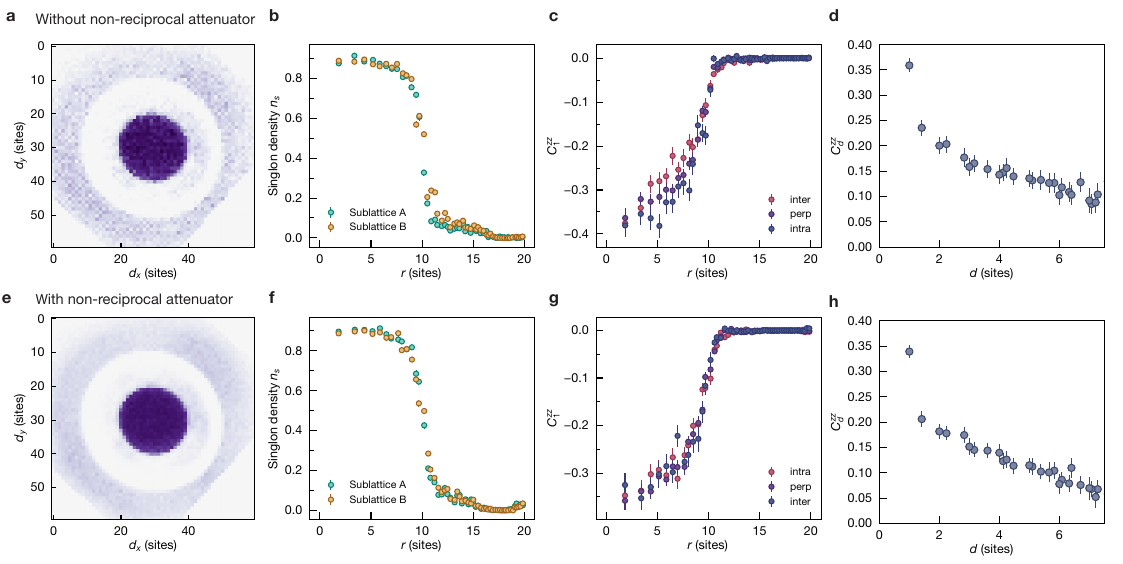"}
    \caption{\textbf{Comparison between experimental methods with and without non-reciprocal attenuator at half-filling.}
    Left to right: averages of two-dimensional singlon density, radially averaged singlon density on A/B sublattices, nearest neighbor spin correlation $C^{zz}(1)$ for intra-dimer, inter-dimer and perpendicular bonds, spin correlations $C^{zz}_d$ averaged over an ROI of $r=5$. The sharp edge in the spatial singlon density profile is due to the edge of the camera sensor.
    \textbf{(a)-(d)}. Without the non-reciprocal attenuator, we ramp the interference time phase to $\phi=\pi/2$ to remove the tunneling dimerization. At half-filling, the resulting sublattice potential offset would not induce a density offset in the Mott insulator region but would cause offsets away from half-filling. The density offset is clearly visible in the outer reservoir region.
    \textbf{(e)-(h)}. With the non-reciprocal attenuator, we could keep the interference phase at $\phi=0$. This suppresses tunneling dimerization below statistical fluctuations without inducing sublattice potential offsets. We find no difference between the densities on the two sublattices. No density modulation is detected in the reservoir.
    }
    \label{sfig:isolator}
\end{dfigure*}

\begin{dfigure*}{qmc_data}
    \centering
    \noindent
    \includegraphics[width=\linewidth]{"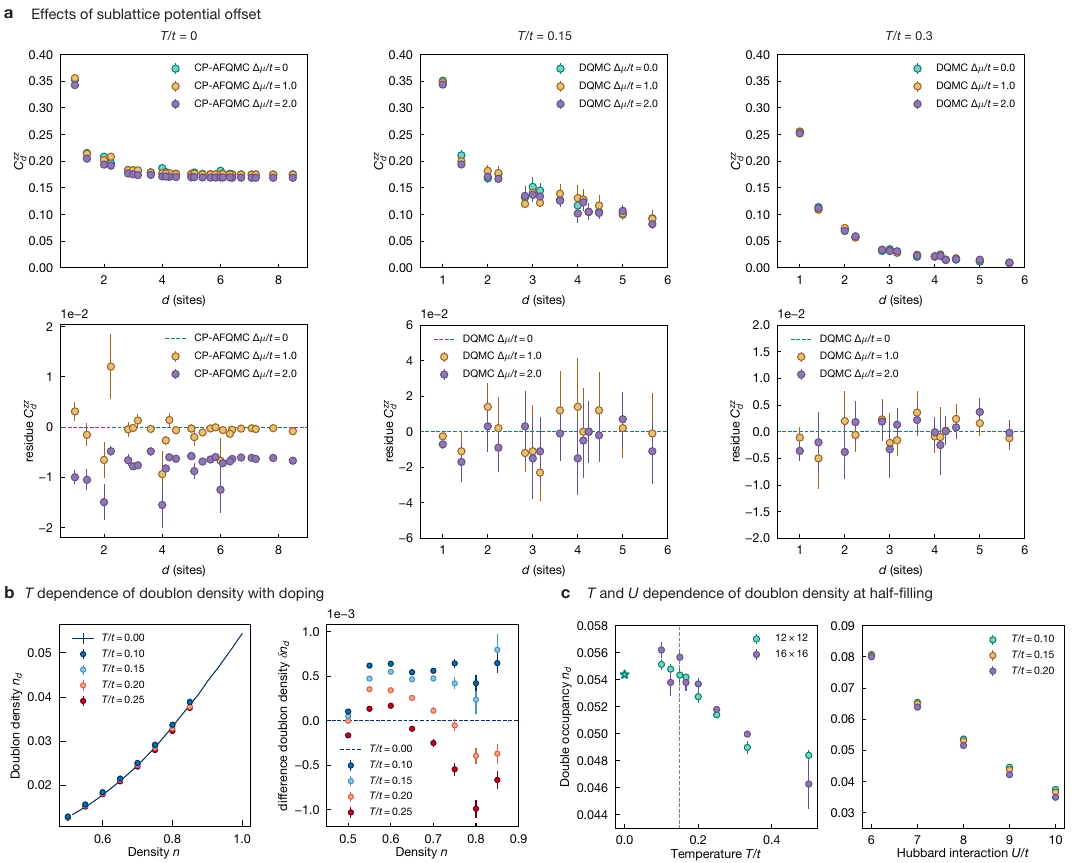"}
    \caption{\textbf{Supplementary data from numerical simulations.}
    \textbf{(a)}. The effects of symmetric A/B sublattice potential offset by $\pm\Delta\mu$ on spin correlations at $T/t=0.15,0.3$ using DQMC and $T/t=0$ using CP-AFQMC, with $\Delta\mu=0.0t, 1.0t, 2.0t$ that is symmetric $\pm \Delta\mu/2$ from $0$ on A/B sublattices . We set the chemical potential $\mu=0$ at half-filling therefore there is no effect on the average density $n=1$. The densities deviation on the A/B sublattices from the average density are $\pm 0, \pm2\%, \pm4\%$ correspondingly. We plot the spin correlations (top) and the difference from the case with no offset (bottom). No statistically significant effect on the spin correlations is detected for $T/t\ge 0.15$ or $T/t=0, \Delta\mu \le 1.0t$. At $T/t=0$, a potential offset of $\Delta\mu=2.0t$ seems to decrease the strength of spin correlations. Note here we use PBC and the long-range AFM order enhanced by finite size effects can be seen as saturation of spin correlation as a function of distance. 
    \textbf{(b)}. (Top) The doublon density $n_d$ as a function of density $n$ computed using CP-AFQMC at different temperatures. Below $T/t=0.15$ the variation of $n_d$ as a function of temperature is small compared to the statistical errors of the detected densities for all dopings where CP-AFQMC simulation is performed. (Bottom) The difference of finite temperature and ground state doublon densities $n_d(T)-n_d(0)$.
    \textbf{(c)}. (Top) The doublon density $n_d$ as a function of temperature $T/t$ at half-filling with interaction $U/t=8$ computed using AFQMC. Below $T/t=0.2$ the variation of $n_d$ as a function of temperature is small compared to the statistical errors of the detected singlon densities. Below $T/t=0.15$ the $n_d$ is consistent with staying constant as it ground state value. (Bottom) The doublon density $n_d$ is sensitive to interaction strengths $U/t$ computed using DQMC for $T/t=0.1,0.15,0.2$. The saturation of $n_d$ as a function of temperature below $T/t=0.15$ holds for a wide range of $U/t$.
    }
    \label{sfig:qmc_data}
\end{dfigure*}

\begin{dfigure*}{exp_qmc}
    \centering
    \noindent
    \includegraphics[width=\linewidth]{"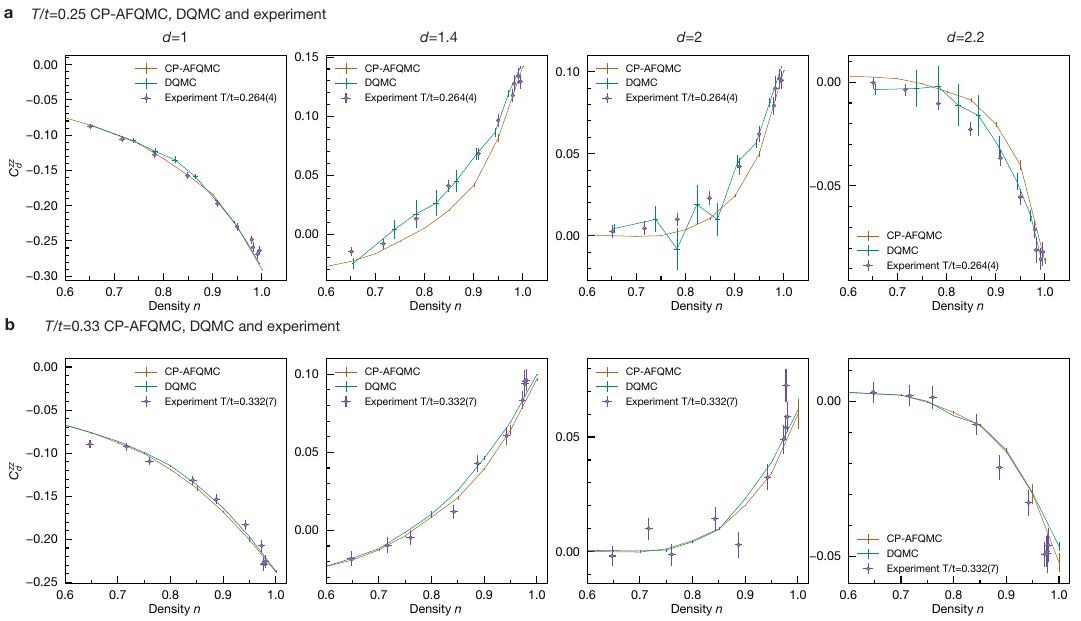"}
    \caption{\textbf{Discrepancy between CP-AFQMC results and experimental data or DQMC results.}
    \textbf{(a)}. Spin correlations computed using DQMC in an $8\times 8$ system and using CP-AFQMC in a $12\times 12$ system at $T/t=0.25$ and obtained from experimental data with a calibrated $U/t=8.2(2), T/t=0.264(4)$.
    \textbf{(b)}. Spin correlations computed using DQMC and using CP-AFQMC in a $12\times 12$ system at $T/t=0.33$ and obtained from experimental data with a calibrated $U/t=8.2(2), T/t=0.332(7)$.
    The experimental data are taken using standard loading with a harmonic trap similar to Ref.~\cite{parsons_site-resolved_2016}. We tune the atom number to reach half-filling in the center of the trap where the atoms form a large Mott insulator that can be used for thermometry. Outside the Mott insulating region, the density decrease as radial distance increases. We obtain spin correlations in the doped region by performing radial binning of the data. The singlon density $n_s$ is converted to density $n$ using DQMC data at $T/t=0.25$.
    }
    \label{sfig:exp_qmc}
\end{dfigure*}

\begin{dfigure*}{experiment_verify}
    \centering
    \noindent
    \includegraphics[width=\linewidth]{"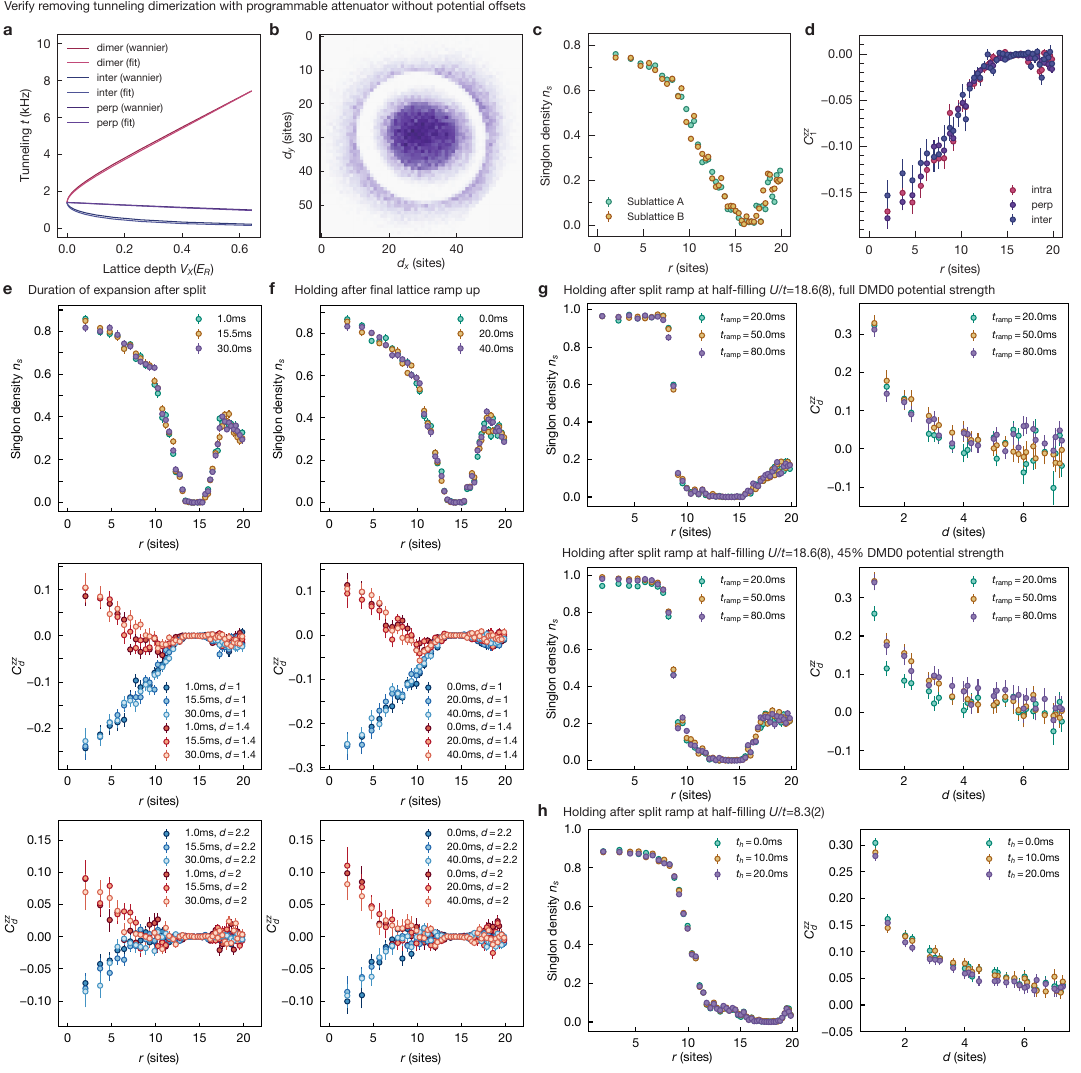"}
    \caption{\textbf{Supplemental experimental data to investigate potential homogeneity and verify the state to be thermalized and equilibrated.}
    \textbf{(a)}. Tunnelings $t_d, t_i, t_p$ as a function of $X$ lattice depth computed by fitting the band structure or constructing wannier orbitals. To balance the tunnelings within $5\%$, we need to decrease $V_X$ below $10^{-4}E_R$. DS7 is used as an example to show the density and correlation profile.
    \textbf{(b)}. The spatial profile of the detected singlon density of the final two dimensional doped Hubbard system in the combined potential defined by the optical lattice and DMDs.
    \textbf{(c)}. Nearest neighbor spin correlation $C^{zz}(1)$ for intra-dimer, inter-dimer and perpendicular bonds. The spin correlations are statistically consistent, confirming the systematical error caused by tunneling dimerization is small compared to statistical error. 
    \textbf{(d)}. Azimuthal average of the singlon density profile shown in \textbf{(b)} on $A/B$ sublattices. We find no difference between the densities on the two sublattices which confirms the absence of sublattice potential offset with $\phi=0$.
    \textbf{(e)}. Density and spin correlations as a function of holding time before the final lattice ramp after expansion. We find no change both in the detected singlon density profile and spin correlation $C^{zz}(|\mathbf{d}|)$ for bond distances $\mathbf{d}=(1,0), (1,1), (2,0), (2,1)$. This confirms $1$\,ms is sufficiently adiabatic for expansion.
    \textbf{(f)}. Density and spin correlations as a function of holding time after the final lattice ramp. This confirms the system is in thermal equilibrium.
    \textbf{(g)}. (Top) At full DMD potential strengths as used to prepare the BI, no statistically significant changes in density or spin correlations are measured as ramp duration is increased from $20$\,ms to $50$\,ms. (Bottom) At reduced DMD potential strengths, fast ramps of $20$\,ms becomes no adiabatic, resulting in significant reduction in density and spin correlations.
    \textbf{(h)}. Density and spin correlations as a function of holding time after the splitting ramp performed at half-filling as in Fig.~\ref{fig:half-filling-hubbard}. The shape of density as a function of radial distances $r$ and spin correlations as a function of bond distance $d$ remain the same after holding suggests the system has thermalized and equilibrated. The reduction of the nearest-neighbor correlation may be due to lattice heating.
    }
    \label{sfig:experiment_verify}
\end{dfigure*}

\begin{dfigure*}{thermal}
    \centering
    \noindent
    \includegraphics[width=\linewidth]{"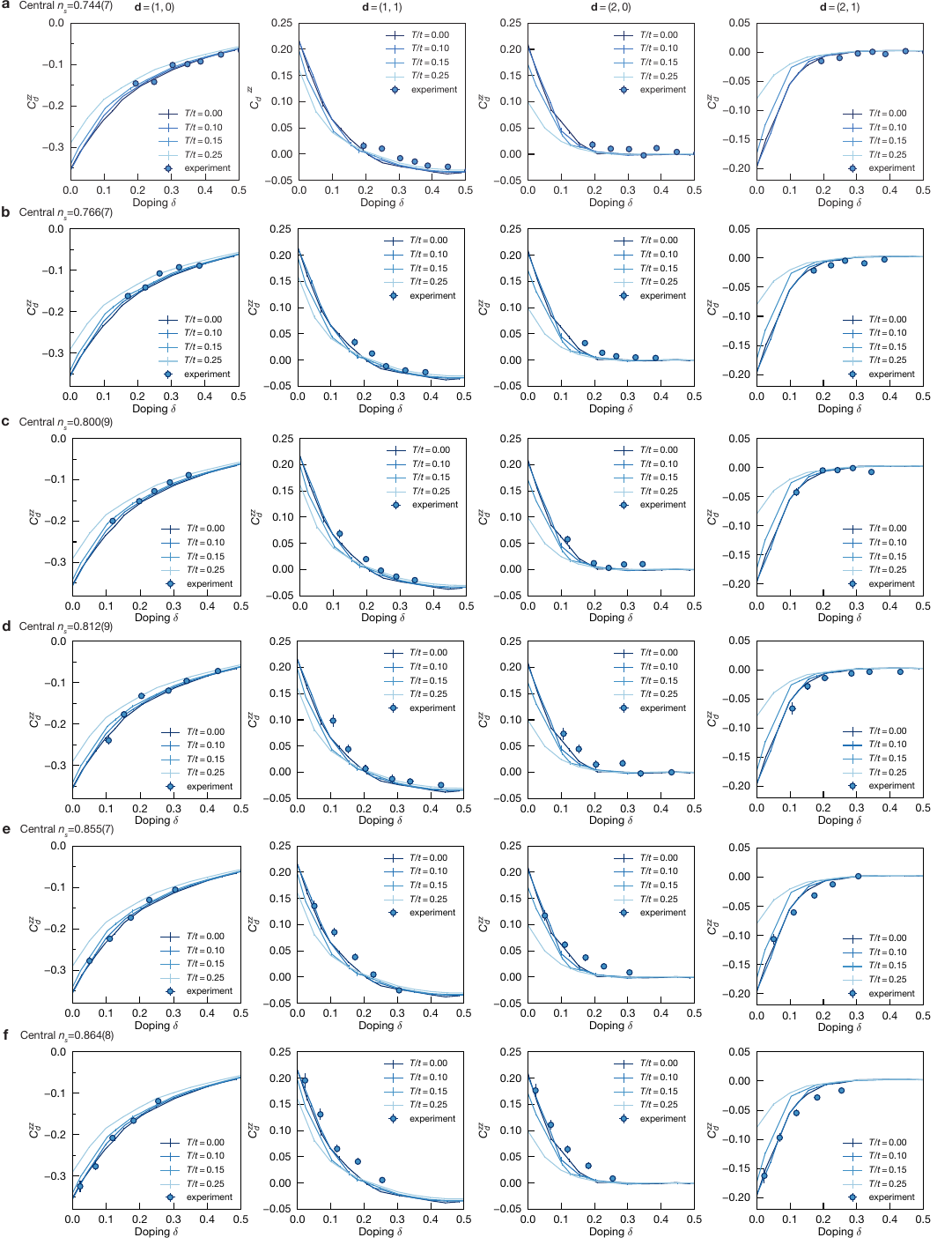"}
    \caption{\textbf{Radially averaged spin correlations vs radially averaged densities.}
    \textbf{(a)-(f)}. Non-strict binned radially averaged spin correlations plotted against the densities of the corresponding bins, for central singlon density $n_s=0.744(7), 0.766(7), 0.800(9), 0.812(9), 0.855(7), 0.864(8)$. Each bin contains $60$ sites, which is the same as an ROI of $r=4$.
    }
    \label{sfig:thermalization}
\end{dfigure*}

\begin{dfigure}{hqmc}
    \centering
    \noindent
    \includegraphics[width=\columnwidth]{"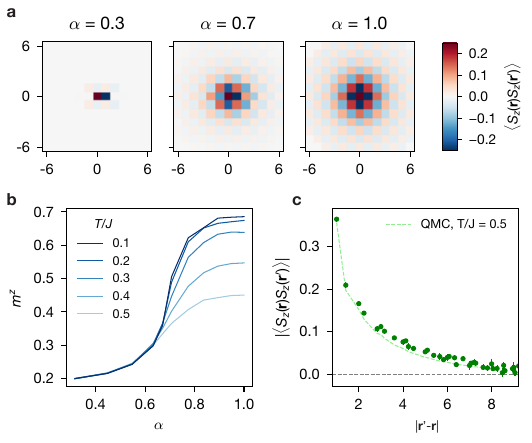"}
    \caption{\textbf{Numerical Quantum Monte Carlo simulation of the Heisenberg model in a dimerized square lattice.}
    \textbf{a}, Bond maps of spin correlation for coupling $\alpha = 0.3, 0.7, 1.0$ at temperature $T/J = 0.5$. \textbf{b}, Staggered magnetization as a function of coupling and temperature. \textbf{c}, Sign-corrected spin correlations as a function of bond distance for experimental data at $U/t = 18.6(8)$ (markers) and fit of square Heisenberg data, yielding a temperature $T/J = 0.458(3)$.}
    \label{sfig:hqmc}
\end{dfigure}

\begin{dfigure}{hf_temperature}
    \centering
    \noindent
    \includegraphics[width=\columnwidth]{"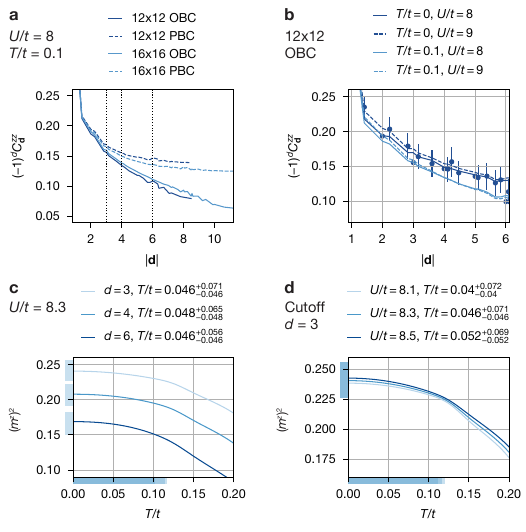"}
    \caption{\textbf{Systematic errors on temperature at half-filling.} \textbf{a,} Effect of boundary conditions on DQMC simulations of the sign-corrected spin correlation function $(-1)^d C_\textbf{d}^{zz}$. \textbf{b,} Effect of interaction strength $U/t$ on simulated $(-1)^d C_\textbf{d}^{zz}$. \textbf{c,} Effect of cutoff bond distance on staggered magnetization square $(m^z)^2$ and estimated temperatures. Shaded rectangles along the y- and x-axis indicate $1\sigma$ confidence intervals on measured $(m^z)^2$ and estimated $T/t$, respectively. \textbf{d,} Effect of uncertainty on calibrated interaction strength on estimated temperatures.
    }
    \label{sfig:hf_temperature}
\end{dfigure}

\end{document}